\documentclass[onecolumn]{article}%
\usepackage{amsmath}
\usepackage{amsfonts}
\usepackage{amssymb}
\usepackage{graphicx}
\usepackage{fullpage}%
\providecommand{\U}[1]{\protect\rule{.1in}{.1in}}
\newtheorem{theorem}{Theorem}[section]

\newtheorem{corollary}[theorem]{Corollary}

\newtheorem{definition}[theorem]{Definition}

\newtheorem{lemma}[theorem]{Lemma}

\newtheorem{objection}{Objection}

\newtheorem{proposition}[theorem]{Proposition}

\newenvironment{proof}[1][Proof]{\noindent\textbf{#1.} }{\ \rule{0.5em}{0.5em}}
\newenvironment{response}[1][Response]{\noindent\textbf{#1.} }{}
\begin{document}

\title{The Learnability of Quantum States}
\author{Scott Aaronson\thanks{Email: scott@scottaaronson.com. \ Supported by CIAR
through the Institute for Quantum Computing.}\\University of Waterloo}
\date{}
\maketitle

\begin{abstract}
Traditional quantum state tomography requires a number of measurements that
grows exponentially with the number of qubits $n$. But using ideas from
computational learning theory, we show that \textquotedblleft for most
practical purposes\textquotedblright\ one can learn a state using a number of
measurements that grows only linearly with $n$. \ Besides possible
implications for experimental physics, our learning theorem has two
applications to quantum computing: first, a new simulation of quantum one-way
communication protocols, and second, the use of trusted classical advice to
verify untrusted quantum advice.

\end{abstract}

\section{Introduction\label{INTRO}}

Suppose we have a physical process that produces a quantum state. \ By
applying the process repeatedly, we can prepare as many copies of the
state\ as we want, and can then measure each copy in a basis of our choice.
\ The goal is to learn an approximate description of the state by combining
the various measurement outcomes.

This problem is called \textit{quantum state tomography},\ and it is already
an important task in experimental physics. \ To give some examples, tomography
has been used to obtain a detailed picture of a chemical reaction (namely, the
dissociation of I$_{2}$ molecules) \cite{ssjm}; to confirm the preparation of
three-photon \cite{rwz} and eight-ion \cite{haffner}\ entangled states; to
test controlled-NOT gates \cite{obrien}; and to characterize optical devices
\cite{dariano}.

Physicists would like to scale up tomography to larger systems, in order to
study the many-particle entangled states that arise (for example)
in\ chemistry, condensed-matter physics, and quantum information. \ But there
is a fundamental obstacle in doing so. \ This is that, to reconstruct an
$n$-qubit state, one needs to measure a number of observables that grows
exponentially in $n$: in particular like $4^{n}$, the number of parameters in
a $2^{n}\times2^{n}$\ density matrix. \ This exponentiality is certainly a
practical problem---H\"{a}ffner et al.\ \cite{haffner}\ report that, to
reconstruct an entangled state of eight calcium ions, they needed to perform
$656,100$ experiments! \ But to us it is a theoretical problem as well. \ For
it suggests that learning an arbitrary state of (say) a thousand particles
would take longer than the age of the universe, even for a being with
unlimited computational power. \ This, in turn, raises the question of what
one even \textit{means} when talking about such a state. \ For whatever else a
quantum state might be, \textit{at the least} it ought to be a hypothesis that
encapsulates previous observations of a physical system, and thereby lets us
predict future observations!

Our purpose here is to propose a new resolution of this conundrum. \ We will
show that, to predict the outcomes of \textquotedblleft most\textquotedblright%
\ measurements on a quantum state, it suffices to do what we call
\textit{pretty-good tomography}---requiring a number of measurements that
grows only \textit{linearly} with the number of qubits $n$.

As a bonus, we will be able to use our learning theorem to prove two new
results in quantum computing and information. \ The first result is a new
relationship between randomized and quantum one-way communication
complexities: namely that $\operatorname*{R}\nolimits^{1}\left(  f\right)
=O\left(  M\operatorname*{Q}\nolimits^{1}\left(  f\right)  \right)  $ for any
partial or total Boolean function $f$, where $M$ is the length of the
recipient's input. \ The second result says that \textit{trusted classical
advice can be used to verify untrusted quantum advice on most inputs}---or in
terms of complexity classes, that $\mathsf{HeurBQP/qpoly}\subseteq
\mathsf{HeurQMA/poly}$. \ Both of these results follow from our learning
theorem in intuitively-appealing ways; on the other hand, we would have no
idea how to prove these results without the theorem. \ To us, this provides
strong evidence that our model for quantum state learning is, if not `the
right one,'\ then certainly \textit{a} right one.

We wish to stress that the main contribution of this paper is
conceptual rather than technical. \ All of the `heavy mathematical
lifting' needed to prove the learning theorem has already been done:
once one has the appropriate setup, the theorem follows readily by
combining previous results due to Bartlett and Long
\cite{bartlettlong} and Ambainis et al.\ \cite{antv}. \ Indeed, what
is surprising to us is precisely that such a basic theorem was not
discovered earlier.

In the remainder of this introduction, we first give a formal statement of our
learning theorem, then answer objections to it, situate it in the context of
earlier work, and discuss its implications.

\subsection{Statement of Result\label{RESULT}}

Let $\rho$\ be an $n$-qubit mixed state: that is, a $2^{n}\times2^{n}%
$\ Hermitian positive semidefinite matrix with $\operatorname*{Tr}\left(
\rho\right)  =1$. \ By a \textit{measurement} of $\rho$, we will mean a
\textquotedblleft two-outcome POVM\textquotedblright: that is, a $2^{n}%
\times2^{n}$\ Hermitian matrix $E$\ with eigenvalues in $\left[  0,1\right]
$. \ Such a measurement $E$ \textit{accepts} $\rho$\ with probability
$\operatorname*{Tr}\left(  E\rho\right)  $, and \textit{rejects} $\rho$\ with
probability $1-\operatorname*{Tr}\left(  E\rho\right)  $.

Our goal will be to learn\ $\rho$. \ Our notion of \textquotedblleft
learning\textquotedblright\ here is purely operational: we want a procedure
that, given a measurement $E$, estimates the acceptance
probability\ $\operatorname*{Tr}\left(  E\rho\right)  $. \ Of course,
estimating $\operatorname*{Tr}\left(  E\rho\right)  $\ for \textit{every} $E$
is the same as estimating $\rho$\ itself, and we know this requires
exponentially many measurements. \ So if we want to learn $\rho$\ using fewer
measurements, then we will have to settle for some weaker success criterion.
\ The criterion we adopt is that we should be able to estimate
$\operatorname*{Tr}\left(  E\rho\right)  $\ for \textit{most}\ measurements
$E$. \ In other words, we assume there is some (possibly unknown) probability
distribution $\mathcal{D}$\ from which the measurements are
drawn.\footnote{$\mathcal{D}$\ can also be a continuous probability measure;
this will not affect any of our results.} \ We are given a \textquotedblleft
training set\textquotedblright\ of measurements\ $E_{1},\ldots,E_{m}$\ drawn
independently from $\mathcal{D}$, as well as the approximate values of
$\operatorname*{Tr}\left(  E_{i}\rho\right)  $ for $i\in\left\{
1,\ldots,m\right\}  $. \ Our goal is to estimate $\operatorname*{Tr}\left(
E\rho\right)  $\ for most $E$'s drawn from $\mathcal{D}$, with high
probability over the choice of training set.

We will show that this can be done using a number of training measurements $m$
that grows only \textit{linearly} with the number of qubits $n$, and
inverse-polynomially with the relevant error parameters. \ Furthermore, the
learning procedure that achieves this bound is the simplest one
imaginable:\ it suffices to find any \textquotedblleft hypothesis
state\textquotedblright\ $\sigma$\ such that $\operatorname*{Tr}\left(
E_{i}\sigma\right)  \approx\operatorname*{Tr}\left(  E_{i}\rho\right)  $ for
all $i$. \ Then with high probability that hypothesis will \textquotedblleft
generalize,\textquotedblright\ in the sense that $\operatorname*{Tr}\left(
E\sigma\right)  \approx\operatorname*{Tr}\left(  E\rho\right)  $ for most
$E$'s drawn from $\mathcal{D}$. \ More precisely:

\begin{theorem}
\label{qoccam}Let $\rho$\ be an $n$-qubit mixed state, let $\mathcal{D}$\ be a
distribution over two-outcome measurements of $\rho$, and let $\mathcal{E}%
=\left(  E_{1},\ldots,E_{m}\right)  $\ be a \textquotedblleft training
set\textquotedblright\ consisting of $m$ measurements drawn independently from
$\mathcal{D}$. \ Also, fix error parameters $\varepsilon,\eta,\gamma
>0$\ with\ $\gamma\varepsilon\geq7\eta$. \ Call $\mathcal{E}$ a
\textquotedblleft good\textquotedblright\ training set if any hypothesis
$\sigma$\ that satisfies%
\[
\left\vert \operatorname*{Tr}\left(  E_{i}\sigma\right)  -\operatorname*{Tr}%
\left(  E_{i}\rho\right)  \right\vert \leq\eta
\]
for all $E_{i}\in\mathcal{E}$, also satisfies%
\[
\Pr_{E\in\mathcal{D}}\left[  \left\vert \operatorname*{Tr}\left(
E\sigma\right)  -\operatorname*{Tr}\left(  E\rho\right)  \right\vert
>\gamma\right]  \leq\varepsilon.
\]
Then there exists a constant $K>0$\ such that $\mathcal{E}$\ is a good
training set with probability at least $1-\delta$, provided that%
\[
m\geq\frac{K}{\gamma^{2}\varepsilon^{2}}\left(  \frac{n}{\gamma^{2}%
\varepsilon^{2}}\log^{2}\frac{1}{\gamma\varepsilon}+\log\frac{1}{\delta
}\right)  .
\]

\end{theorem}

\subsection{Objections and Variations\label{OBJECTIONS}}

Before proceeding further, it will be helpful to answer various objections
that might be raised against Theorem \ref{qoccam}. \ Along the way, we will
also state two variations of the theorem.\medskip

\begin{objection}
By changing the goal to \textquotedblleft pretty
good\ tomography,\textquotedblright\ Theorem \ref{qoccam}\ dodges much of the
quantum state tomography problem as ordinarily understood.
\end{objection}

\begin{response}
Yes, that is exactly what it does! \ The motivating idea is that one does not
need to know the expectation values for \textit{all} observables, only for
most of the observables that will actually be measured. \ As an example, if we
can only apply $1$- and $2$-qubit measurements, then the outcomes of $3$-qubit
measurements are irrelevant by assumption. \ As a less trivial example,
suppose the measurement distribution $\mathcal{D}$ is uniformly random (i.e.,
is the Haar measure). \ Then even if our quantum system is \textquotedblleft
really\textquotedblright\ in some pure state $\left\vert \psi\right\rangle $,
for reasonably large $n$ it will be billions of years before we happen upon a
measurement that distinguishes $\left\vert \psi\right\rangle $ from the
maximally mixed state. \ Hence the maximally mixed state is perfectly adequate
as an explanatory hypothesis, despite being far from $\left\vert
\psi\right\rangle $ in the usual metrics such as trace distance.

Of course, even after one relaxes the goal in this way, it might still seem
surprising that for any state $\rho$, and any distribution $\mathcal{D}$, a
linear amount of tomographic data is sufficient to simulate most measurements
drawn from $\mathcal{D}$. \ This is the content of Theorem \ref{qoccam}.
\end{response}

\bigskip

\begin{objection}
But to apply Theorem \ref{qoccam}, one needs the measurements to be drawn
independently from some probability distribution $\mathcal{D}$. \ Is this not
a strange assumption? \ Shouldn't one also allow adaptive measurements?
\end{objection}

\begin{response}
If all of our training data involved measurements in the $\left\{  \left\vert
0\right\rangle ,\left\vert 1\right\rangle \right\}  $ basis, then regardless
of how much data we had, clearly we couldn't hope to simulate a measurement in
the $\left\{  \left\vert +\right\rangle ,\left\vert -\right\rangle \right\}
$\ basis! \ Therefore, as usual in learning theory, to get anywhere we need to
make \textit{some} assumption to the effect that the future will resemble the
past. \ Such an assumption does not strike us as unreasonable in the context
of quantum state estimation. \ For example, suppose that (as is often the
case) the measurement process was itself stochastic, so that the experimenter
did not know which observable was going to be measured until after it
\textit{was} measured. \ Or suppose the state was a \textquotedblleft quantum
program,\textquotedblright\ which only had to succeed on typical inputs drawn
from some probability distribution.\footnote{At this point we should remind
the reader that the distribution $\mathcal{D}$\ over measurements only has to
\textit{exist}; it does not have to be \textit{known}. \ All of our learning
algorithms will be \textquotedblleft distribution-free,\textquotedblright\ in
the sense that a single algorithm will work for any choice of $\mathcal{D}$.}

However, with regard to the power of adaptive measurements, it is possible to
ask slightly more sophisticated questions. \ For example, suppose we perform a
binary measurement $E_{1}$\ (drawn from some distribution $\mathcal{D}$) on
one copy of an $n$-qubit state $\rho$. \ Then, based on the outcome $z_{1}%
\in\left\{  0,1\right\}  $\ of that measurement, suppose we perform another
binary measurement $E_{2}$\ (drawn from a new distribution $\mathcal{D}%
_{z_{1}}$) on a second copy of $\rho$; and so on for $r$ copies of $\rho$.
\ Finally, suppose we compute some Boolean function $f\left(  z_{1}%
,\ldots,z_{r}\right)  $\ of the $r$ measurement outcomes.

Now, how many times will we need to repeat this adaptive procedure before,
given $E_{1},\ldots,E_{r}$ drawn as above, we can estimate (with high
probability) the conditional probability that $f\left(  z_{1},\ldots
,z_{r}\right)  =1$? \ If we simply apply Theorem \ref{qoccam}\ to the tensor
product of all $r$ registers, then it is easy to see that $O\left(  nr\right)
$\ samples suffice. \ Furthermore, using the ideas of Appendix \ref{LB},\ one
can show that this is optimal: in other words, no improvement to (say)
$O\left(  n+r\right)  $\ samples is possible.

Indeed, even if we wanted to estimate the probabilities of all $r$ of the
measurement outcomes \textit{simultaneously}, it follows from the union bound
that we could do this with high probability, after a number of samples linear
in $n$ and polynomial in $r$.

We hope this illustrates how our learning theorem\ can be applied to more
general settings than that for which it is explicitly stated. \ Naturally,
there is a great deal of scope here for further research.
\end{response}

\bigskip

\begin{objection}
Theorem \ref{qoccam} is purely information-theoretic;\ as such, it says
nothing about the \textit{computational} complexity of finding a hypothesis
state\ $\sigma$.
\end{objection}

\begin{response}
This is correct. \ Using semidefinite and convex programming techniques, one
can implement any of our learning algorithms to run in time polynomial in the
Hilbert space dimension, $N=2^{n}$. \ This might be fine if $n$ is at most
$12$ or so; note that \textquotedblleft measurement
complexity,\textquotedblright\ and not computational complexity, has almost
always been the limiting factor in real experiments. \ But of course such a
running time is prohibitive for larger $n$.

Let us stress that exactly the same problem arises even in classical learning
theory. \ For it follows from a celebrated result of Goldreich, Goldwasser,
and Micali \cite{ggm}\ that, if there exists a polynomial-time algorithm to
find a Boolean circuit of size $n$\ consistent with observed data (whenever
such a circuit exists), then there are no cryptographic one-way functions.
\ Using the same techniques, one can show that, if there exists a
polynomial-time quantum algorithm to prepare a state of $n^{k}$\ qubits
consistent with observed data (whenever such a state exists), then there are
no (classical) one-way functions secure against quantum attack. \ The only
difference is that, while finding a classical hypothesis consistent with data
is an $\mathsf{NP}$ search problem,\footnote{Interestingly, in the
\textquotedblleft representation-independent\textquotedblright\ setting (where
the output hypothesis can be an arbitrary Boolean circuit), this problem is
\textit{not} known to be $\mathsf{NP}$-complete.} finding a quantum hypothesis
is a $\mathsf{QMA}$ search problem.

A fundamental question left open by this paper is whether there are nontrivial
special cases of the quantum learning problem that can be solved, not only
with a linear number of measurements, but also with a polynomial amount of
quantum computation.
\end{response}

\bigskip

\begin{objection}
The dependence on the error parameters $\gamma$ and $\varepsilon$ in Theorem
\ref{qoccam}\ looks terrible.
\end{objection}

\begin{response}
Indeed, no one would pretend that performing $\sim\frac{1}{\gamma
^{4}\varepsilon^{4}}$ measurements is practical for reasonable $\gamma$ and
$\varepsilon$. \ Fortunately, we can improve the dependence on $\gamma$\ and
$\varepsilon$\ quite substantially, at the cost of increasing the dependence
on $n$ from linear to $n\log^{2}n$.

\noindent

\begin{theorem}
\label{qoccam2}The bound in Theorem \ref{qoccam} can be replaced by%
\[
m\geq\frac{K}{\varepsilon}\left(  \frac{n}{\left(  \gamma-\eta\right)  ^{2}%
}\log^{2}\frac{n}{\left(  \gamma-\eta\right)  \varepsilon}+\log\frac{1}%
{\delta}\right)
\]
for all $\varepsilon,\eta,\gamma>0$\ with\ $\gamma>\eta$.
\end{theorem}

In Appendix \ref{LB}, we will show that the dependence on $\gamma$ and
$\varepsilon$\ in Theorem \ref{qoccam2} is close to optimal.
\end{response}

\bigskip

\begin{objection}
To estimate the measurement probabilities $\operatorname*{Tr}\left(  E_{i}%
\rho\right)  $, one needs the ability to prepare multiple copies of $\rho$.
\end{objection}

\begin{response}
This is less an objection to Theorem \ref{qoccam}\ than to quantum mechanics
itself! \ If one has only one copy of $\rho$, then Holevo's Theorem
\cite{holevo} immediately implies that not even \textquotedblleft pretty good
tomography\textquotedblright\ is possible.
\end{response}

\bigskip

\begin{objection}
Even with unlimited copies of $\rho$, one could never be certain that the
condition of Theorem \ref{qoccam} was satisfied (i.e., that $\left\vert
\operatorname*{Tr}\left(  E_{i}\sigma\right)  -\operatorname*{Tr}\left(
E_{i}\rho\right)  \right\vert \leq\eta$ for every $i$).
\end{objection}

\begin{response}
This is correct, but there is no need for certainty. \ For suppose we apply
each measurement $E_{i}$\ to $\Theta\left(  \frac{\log m}{\eta^{2}}\right)  $
copies of $\rho$. \ Then by a large deviation bound, with overwhelming
probability we will obtain real numbers $p_{1},\ldots,p_{m}$\ such that
$\left\vert p_{i}-\operatorname*{Tr}\left(  E_{i}\rho\right)  \right\vert
\leq\eta/2$\ for every $i$. \ So if we want to find a hypothesis state
$\sigma$\ such that $\left\vert \operatorname*{Tr}\left(  E_{i}\sigma\right)
-\operatorname*{Tr}\left(  E_{i}\rho\right)  \right\vert \leq\eta$\ for every
$i$, then it suffices to find a $\sigma$\ such that $\left\vert p_{i}%
-\operatorname*{Tr}\left(  E_{i}\sigma\right)  \right\vert \leq\eta/2$\ for
every $i$. \ Certainly such a $\sigma$\ exists, for take $\sigma=\rho$.
\end{response}

\bigskip

\begin{objection}
But what if one can apply each measurement only \textit{once}, rather than
multiple times? \ In that case,\ the above estimation strategy no longer works.
\end{objection}

\begin{response}
In Appendix \ref{MEAS}, we will prove a learning theorem that applies directly
to this \textquotedblleft measure-once\textquotedblright\ scenario. \ The
disadvantage is that the upper bound on the number of measurements will
increase from $\sim1/\left(  \gamma^{4}\varepsilon^{4}\right)  $\ to
$\sim1/\left(  \gamma^{8}\varepsilon^{4}\right)  $.

\noindent

\begin{theorem}
\label{qoccam3}Let $\rho$ be an $n$-qubit state, let $\mathcal{D}$ be a
distribution over two-outcome measurements, and let $\mathcal{E}=\left(
E_{1},\ldots,E_{m}\right)  $\ consist of $m$ measurements drawn independently
from $\mathcal{D}$. \ Suppose we are given bits $B=\left(  b_{1},\ldots
,b_{m}\right)  $, where each $b_{i}$\ is $1$ with independent probability
$\operatorname*{Tr}\left(  E_{i}\rho\right)  $ and $0$\ with probability
$1-\operatorname*{Tr}\left(  E_{i}\rho\right)  $. \ Suppose also that we
choose a hypothesis state $\sigma$\ to minimize the quadratic functional
$\sum_{i=1}^{m}\left(  \operatorname*{Tr}\left(  E_{i}\sigma\right)
-b_{i}\right)  ^{2}$. \ Then there exists a positive constant $K$ such that%
\[
\Pr_{E\in\mathcal{D}}\left[  \left\vert \operatorname*{Tr}\left(
E\sigma\right)  -\operatorname*{Tr}\left(  E\rho\right)  \right\vert
>\gamma\right]  \leq\varepsilon
\]
with probability at least $1-\delta$\ over $\mathcal{E}$\ and $B$, provided
that%
\[
m\geq\frac{K}{\gamma^{4}\varepsilon^{2}}\left(  \frac{n}{\gamma^{4}%
\varepsilon^{2}}\log^{2}\frac{1}{\gamma\varepsilon}+\log\frac{1}{\delta
}\right)  .
\]

\end{theorem}
\end{response}

\bigskip

\begin{objection}
What if, instead of applying the \textquotedblleft ideal\textquotedblright%
\ measurement $E$, the experimenter can only apply a noisy version $E^{\prime
}$?
\end{objection}

\begin{response}
If the noise that corrupts $E$\ to $E^{\prime}$\ is governed by a known
probability distribution such as a Gaussian, then $E^{\prime}$ is still just a
POVM, so Theorem \ref{qoccam}\ applies directly. \ If the noise is
adversarial, then we can also apply Theorem \ref{qoccam} directly, provided we
have an upper bound on $\left\vert \operatorname*{Tr}\left(  E^{\prime}%
\rho\right)  -\operatorname*{Tr}\left(  E\rho\right)  \right\vert $ (which
simply gets absorbed into $\eta$).
\end{response}

\bigskip

\begin{objection}
What if the measurements have $k>2$ possible outcomes?
\end{objection}

\begin{response}
Here is a simple reduction to the two-outcome\ case. \ Before applying the
$k$-outcome POVM $E=\left\{  E^{\left(  1\right)  },\ldots,E^{\left(
k\right)  }\right\}  $, first choose an integer $j\in\left\{  1,\ldots
,k\right\}  $\ uniformly at random, and then pretend that the POVM being
applied is $\left\{  E^{\left(  j\right)  },I-E^{\left(  j\right)  }\right\}
$ (i.e., ignore the other $k-1$\ outcomes). \ By the union bound, if our goal
is to ensure that%
\[
\Pr_{E\in\mathcal{D}}\left[  \sum_{j=1}^{k}\left\vert \operatorname*{Tr}%
\left(  E^{\left(  j\right)  }\sigma\right)  -\operatorname*{Tr}\left(
E^{\left(  j\right)  }\rho\right)  \right\vert >\gamma\right]  \leq\varepsilon
\]
with probability at least $1-\delta$, then in our upper bounds it suffices to
replace every occurrence of $\gamma$\ by $\gamma/k$, and every occurrence of
$\varepsilon$\ by $\varepsilon/k$. \ We believe that one could do better than
this by analyzing the $k$-outcome case directly; we leave this as an open
problem.\footnote{Notice that any sample complexity bound must have at least a
linear dependence on $k$. \ Here is a proof sketch: given a subset
$S\subseteq\left\{  1,\ldots,k\right\}  $ with $\left\vert S\right\vert =k/2$,
let $\left\vert S\right\rangle $\ be a uniform superposition over the elements
of $S$. \ Now consider simulating a measurement of $\left\vert S\right\rangle
$\ in the computational basis, $\left\{  \left\vert 1\right\rangle
,\ldots,\left\vert k\right\rangle \right\}  $. \ It is clear that
$\Omega\left(  k\right)  $\ sample measurements are needed to do this even
approximately.}
\end{response}

\subsection{Related Work\label{RELATED}}

This paper builds on two research areas---computational learning theory and
quantum information theory---in order to say something about a third area:
quantum state estimation. \ Since many readers are probably unfamiliar with at
least one of these areas, let us discuss them in turn.\bigskip

\textbf{Computational Learning Theory}

Computational learning theory can be understood as a modern response to David
Hume's Problem of Induction: \textquotedblleft if an ornithologist sees $500$
ravens and all of them are black, why does that provide any grounds at all for
expecting the $501^{st}$ raven to be black?\ \ After all, the hypothesis that
the $501^{st}$ raven will be white seems equally compatible with
evidence.\textquotedblright\ \ The answer, from a learning theory perspective,
is that in practice one always restricts attention to some class $\mathcal{C}%
$\ of hypotheses that is vastly smaller than the class of logically
conceivable hypotheses. \ So the real question is not \textquotedblleft is
induction possible?,\textquotedblright\ but rather \textquotedblleft what
properties does the class $\mathcal{C}$\ have to satisfy for induction to be
possible?\textquotedblright

In a seminal 1989 paper, Blumer et al.\ \cite{behw}\ showed that if
$\mathcal{C}$ is finite, then any hypothesis that agrees with
$O\left( \log\left\vert \mathcal{C}\right\vert \right)  $\
randomly-chosen data points will probably agree with most future
data points as well. \ Indeed, even if $\mathcal{C}$ is infinite,
one can upper-bound the number of data points
needed for learning in terms of a combinatorial parameter\ of $\mathcal{C}%
$\ called the VC (Vapnik-Chervonenkis) dimension. \ Unfortunately, these
results apply only to Boolean hypothesis classes. \ So to prove our learning
theorem, we will need a more powerful result due to Bartlett and Long
\cite{bartlettlong},\ which upper-bounds the number of data points needed to
learn \textit{real}-valued hypothesis classes.\bigskip

\textbf{Quantum Information Theory}

Besides results from classical learning theory, we will also need a
result of Ambainis et al.\ \cite{antv} in quantum information
theory. \ Ambainis et al.\ showed that, if we want to encode $k$
bits into an $n$-qubit quantum state, in such a way that any one
bit\ can later be retrieved with error probability at most $p$, then
we need $n\geq\left(  1-H\left(  p\right)  \right)  k$, where $H$\
is the binary entropy function.

Perhaps the central idea of this paper is to turn Ambainis et al.'s result on
its head, and see it not as lower-bounding the number of qubits needed for
coding and communication tasks, but instead as \textit{upper}-bounding the
\textquotedblleft effective dimension\textquotedblright\ of a quantum state to
be learned. \ (In theoretical computer science, this is hardly the first time
that a negative result has been turned into a positive one. \ A similar
\textquotedblleft lemons-into-lemonade\textquotedblright\ conceptual shift was
made by Linial, Mansour, and Nisan \cite{lmn}, when they used a limitation of
constant-depth circuits to give an efficient algorithm for learning those
circuits.\bigskip)

\textbf{Quantum State Estimation}

Physicists have been interested in quantum state estimation since at least the
1950's (see \cite{qse} for a good overview). \ For practical reasons, they
have been particularly concerned with minimizing the number of measurements.
\ However, most literature on the subject restricts attention to
low-dimensional Hilbert spaces (say, $2$ or $3$ qubits),\ taking for granted
that the number of measurements will increase exponentially with the number of qubits.

There \textit{is} a substantial body of work on how to estimate a quantum
state given incomplete measurement results---see Bu\v{z}ek et
al.\ \cite{bddaw} for a good introduction to the subject, or Bu\v{z}ek
\cite{buzek} for estimation algorithms that are similar in spirit to ours.
\ But there are at least two differences between the previous work and ours.
\ First, while some of the previous work offers numerical evidence that few
measurements seem to suffice in practice, so far as we know none of it
considers asymptotic complexity. \ Second, the previous work almost always
assumes that an experimenter starts with a prior probability distribution over
quantum states (often the uniform distribution), and then either updates the
distribution using Bayes' rule, or else applies a Maximum-Likelihood
principle. \ By contrast, our learning approach requires no assumptions about
a distribution over states; it instead requires only a (possibly-unknown)
distribution over \textit{measurements}. \ The advantage of the latter
approach, in our view, is that an experimenter has much more control over
which measurements to apply than over the nature of the state to be learned. \

\subsection{Implications\label{IMP}}

The main implication of our learning theorem is conceptual: it shows that
quantum states, considered as a hypothesis class, are \textquotedblleft
reasonable\textquotedblright\ in the sense of computational learning theory.
\ Were this \textit{not} the case, it would presumably strengthen the view of
quantum computing skeptics \cite{goldreich:qc,levin:qc} that quantum states
are \textquotedblleft inherently extravagant\textquotedblright\ objects, which
will need to be discarded as our knowledge of physics expands.\footnote{Or at
least, it would suggest that the \textquotedblleft operationally
meaningful\textquotedblright\ quantum states comprise only a tiny portion of
Hilbert space.} \ Instead we have shown that, while the \textquotedblleft
effective dimension\textquotedblright\ of an $n$-qubit Hilbert space
\textit{appears} to be exponential in $n$, in the sense that is relevant for
approximate learning and prediction this appearance is illusory.

Beyond establishing this conceptual point, we believe our learning
theorem\ could be of practical use in quantum state estimation, since it
provides an explicit upper bound on the number of measurements needed to
\textquotedblleft learn\textquotedblright\ a quantum state with respect to any
probability measure over observables.\ \ Even if our actual result is not
directly applicable, we hope the mere \textit{fact} that this sort of
learning\ is possible will serve as a spur to further research. \ As an
analogy, classical computational learning theory has had a large influence on
neural networks, computer vision, and other fields,\footnote{According to
Google Scholar, Valiant's original paper on the subject \cite{valiant:pac} has
been cited 1829 times, with a large fraction of the citations coming from
practitioners.} but this influence might have had less to do with the results
themselves than with their philosophical moral.

We turn now to a more immediate application of our learning theorem: solving
open problems in quantum computing and information.

The first problem concerns \textit{quantum one-way communication complexity}.
\ In this subject we consider a sender, Alice, and a receiver, Bob, who hold
inputs $x$ and $y$ respectively. \ We then ask the following question:
assuming the best communication protocol and the worst $\left(  x,y\right)
$\ pair, how many bits must Alice send to Bob, for Bob to be able to evaluate
some joint function $f\left(  x,y\right)  $ with high probability? \ Note that
there is no back-communication from Bob to Alice.

Let $\operatorname*{R}\nolimits^{1}\left(  f\right)  $, and $\operatorname*{Q}%
\nolimits^{1}\left(  f\right)  $\ be the number of bits that Alice needs to
send, if her message to Bob is randomized or quantum
respectively.\footnote{Here the superscript `$1$' denotes one-way
communication.} \ Then improving an earlier result of Aaronson \cite{aar:adv},
in Section \ref{COMM} we are able to show the following:

\begin{theorem}
\label{ccthm}For any Boolean function $f$ (partial or total),
$\operatorname*{R}\nolimits^{1}\left(  f\right)  =O\left(  M\operatorname*{Q}%
\nolimits^{1}\left(  f\right)  \right)  $, where $M$\ is the length of Bob's input.
\end{theorem}

Intuitively, this means that if Bob's input is small, then quantum
communication provides at most a small advantage over classical communication.

The proof of Theorem \ref{ccthm} will rely on our learning theorem in an
intuitively appealing way. \ Basically, Alice will send some randomly-chosen
\textquotedblleft training inputs,\textquotedblright\ which Bob will then use
to learn a \textquotedblleft pretty good description\textquotedblright\ of the
quantum state that Alice would have sent him in the quantum protocol.

The second problem concerns \textit{approximate verification of quantum
software}. \ Suppose you want to evaluate some Boolean function $f:\left\{
0,1\right\}  ^{n}\rightarrow\left\{  0,1\right\}  $, on typical inputs $x$
drawn from a probability distribution $\mathcal{D}$. \ So you go to the
quantum software store and purchase $\left\vert \psi_{f}\right\rangle $, a
$q$-qubit piece of quantum software. \ The software vendor tells you that, to
evaluate $f\left(  x\right)  $\ on any given input $x\in\left\{  0,1\right\}
^{n}$, you simply need to apply a fixed measurement $E$ to the state
$\left\vert \psi_{f}\right\rangle \left\vert x\right\rangle $. \ However, you
do not trust $\left\vert \psi_{f}\right\rangle $ to work as expected. \ Thus,
the following question arises: is there a fixed, polynomial-size set of
\textquotedblleft benchmark inputs\textquotedblright\ $x_{1},\ldots,x_{T}$,
such that for \textit{any} quantum program $\left\vert \psi_{f}\right\rangle
$, if $\left\vert \psi_{f}\right\rangle $\ works on the benchmark inputs then
it will also work on most inputs drawn from $\mathcal{D}$?

Using our learning theorem, we will show in Appendix \ref{ADV}\ that the
answer is yes. \ Indeed, we will actually go further than that, and give an
\textit{efficient procedure} to test $\left\vert \psi_{f}\right\rangle
$\ against the benchmark inputs. \ The central difficulty here is that the
measurements intended to test $\left\vert \psi_{f}\right\rangle $ might also
destroy it. \ We will resolve this difficulty by means of a \textquotedblleft
Witness Protection Lemma,\textquotedblright\ which might have applications elsewhere.

In terms of complexity classes, we can state our verification theorem as follows:

\begin{theorem}
\label{advthm}$\mathsf{HeurBQP/qpoly}\subseteq\mathsf{HeurQMA/poly}$.
\end{theorem}

Here $\mathsf{BQP/qpoly}$ is the class of problems solvable in quantum
polynomial time, with help from a polynomial-size \textquotedblleft quantum
advice state\textquotedblright\ $\left\vert \psi_{n}\right\rangle $\ that
depends only on the input length $n$; while $\mathsf{QMA}$\ (Quantum
Merlin-Arthur) is the class of problems for which a `yes' answer admits a
polynomial-size quantum proof. \ Then $\mathsf{HeurBQP/qpoly}$\ and
$\mathsf{HeurQMA/poly}$\ are the \textit{heuristic} versions of
$\mathsf{BQP/qpoly}$\ and $\mathsf{QMA/poly}$ respectively---that is, the
versions where we only want to succeed on most inputs rather than all of them.

\section{The Measurement Complexity of Quantum Learning\label{SC}}

We now prove Theorems \ref{qoccam} and \ref{qoccam2}. \ To do so, we
first review results from computational learning theory, which
upper-bound the number of data points needed to learn a hypothesis
in terms of the \textquotedblleft dimension\textquotedblright\ of
the underlying hypothesis class. \ We then use a result of Ambainis
et al.\ \cite{antv}\ to upper-bound the dimension of the class of
$n$-qubit mixed states.

\subsection{Learning Probabilistic Concepts\label{PCONCEPT}}

The prototype of the sort of learning theory result we need is the
\textquotedblleft Occam's Razor Theorem\textquotedblright\ of Blumer et
al.\ \cite{behw}, which is stated in terms of a parameter called VC dimension.
\ However, Blumer et al.'s result\ does not suffice for our purpose, since it
deals with \textit{Boolean} concepts, which map each element of an underlying
sample space to $\left\{  0,1\right\}  $. \ By contrast, we are interested in
probabilistic concepts---called \textit{p-concepts} by Kearns and Schapire
\cite{ks:pconcept}---which map each measurement $E$\ to a real number
$\operatorname*{Tr}\left(  E\rho\right)  \in\left[  0,1\right]  $.

Generalizing from Boolean concepts to p-concepts is not as straightforward as
one might hope. \ Fortunately, various authors
\cite{abch,ab:learn,bartlettlong,blw,ks:pconcept} have already done most of
the work for us,\ with results due to\ Anthony and Bartlett \cite{ab:learn}
and to Bartlett and Long \cite{bartlettlong} being particularly relevant. \ To
state their results, we need some definitions. \ Let $\mathcal{S}$ be a finite
or infinite set called the \textit{sample space}. \ Then a \textit{p-concept
over} $\mathcal{S}$\ is a function $F:\mathcal{S}\rightarrow\left[
0,1\right]  $, and\ a \textit{p-concept class over}\ $\mathcal{S}$\ is a set
of p-concepts over $\mathcal{S}$. \ Kearns and Schapire \cite{ks:pconcept}%
\ proposed a measure of the complexity of p-concept classes, called the
\textit{fat-shattering dimension}.

\begin{definition}
Let $\mathcal{S}$\ be a sample space, let $\mathcal{C}$ be a p-concept class
over $\mathcal{S}$, and let $\gamma>0$ be a real number. \ We say a set
$\left\{  s_{1},\ldots,s_{k}\right\}  \subseteq\mathcal{S}$\ is $\gamma
$-fat-\textit{shattered} by $\mathcal{C}$ if there exist real numbers
$\alpha_{1},\ldots,\alpha_{k}$ such that for all $B\subseteq\left\{
1,\ldots,k\right\}  $, there exists a p-concept $F\in\mathcal{C}$\ such that
for all $i\in\left\{  1,\ldots,k\right\}  $,

\begin{enumerate}
\item[(i)] if $i\notin B$ then $F\left(  s_{i}\right)  \leq\alpha_{i}-\gamma$, and

\item[(ii)] if $i\in B$ then $F\left(  s_{i}\right)  \geq\alpha_{i}+\gamma$.
\end{enumerate}

Then the $\gamma$-fat-\textit{shattering dimension} of $\mathcal{C}$, or
$\operatorname*{fat}\nolimits_{\mathcal{C}}\left(  \gamma\right)  $, is the
maximum $k$ such that some $\left\{  s_{1},\ldots,s_{k}\right\}
\subseteq\mathcal{S}$ is $\gamma$-fat-shattered by $\mathcal{C}$. \ (If there
is no finite such maximum, then $\operatorname*{fat}\nolimits_{\mathcal{C}%
}\left(  \gamma\right)  =\infty$.)
\end{definition}

We can now state the result of Anthony and Bartlett.

\begin{theorem}
[Anthony and Bartlett \cite{ab:learn}]\label{occam}Let $\mathcal{S}$\ be a
sample space, let $\mathcal{C}$\ be a p-concept class over $\mathcal{S}$, and
let $\mathcal{D}$ be a probability measure over $\mathcal{S}$. \ Fix an
element $F\in\mathcal{C}$, as well as error parameters $\varepsilon
,\eta,\gamma>0$\ with\ $\gamma>\eta$.\ \ Suppose we draw $m$
samples\ $X=\left(  x_{1},\ldots,x_{m}\right)  $\ independently according to
$\mathcal{D}$, and then choose any hypothesis $H\in\mathcal{C}$\ such that
$\left\vert H\left(  x\right)  -F\left(  x\right)  \right\vert \leq\eta$\ for
all $x\in X$. \ Then there exists a positive constant $K$ such that%
\[
\Pr_{x\in\mathcal{D}}\left[  \left\vert H\left(  x\right)  -F\left(  x\right)
\right\vert >\gamma\right]  \leq\varepsilon
\]
with probability at least $1-\delta$\ over $X$, provided that%
\[
m\geq\frac{K}{\varepsilon}\left(  \operatorname*{fat}\nolimits_{\mathcal{C}%
}\left(  \frac{\gamma-\eta}{8}\right)  \log^{2}\left(  \frac
{\operatorname*{fat}\nolimits_{\mathcal{C}}\left(  \frac{\gamma-\eta}%
{8}\right)  }{\left(  \gamma-\eta\right)  \varepsilon}\right)  +\log\frac
{1}{\delta}\right)  .
\]

\end{theorem}

Notice that in Theorem \ref{occam}, the dependence on the fat-shattering
dimension\ is superlinear. \ We would like to reduce the dependence to linear,
at least when $\eta$\ is sufficiently small. \ We can do so using the
following result of Bartlett and Long.\footnote{The result we state is a
special case of Bartlett and Long's Theorem 20, where the function $F$ to be
learned is itself a member of the hypothesis class $\mathcal{C}$.}

\begin{theorem}
[Bartlett and Long \cite{bartlettlong}]\label{occam2}Let $\mathcal{S}$\ be a
sample space, let $\mathcal{C}$\ be a p-concept class over $\mathcal{S}$, and
let $\mathcal{D}$ be a probability measure over $\mathcal{S}$. \ Fix a
p-concept $F:\mathcal{S}\rightarrow\left[  0,1\right]  $\ (not necessarily in
$\mathcal{C}$),\ as well as an error parameter $\alpha>0$. \ Suppose we draw
$m$ samples $X=\left(  x_{1},\ldots,x_{m}\right)  $\ independently according
to $\mathcal{D}$, and then choose any hypothesis $H\in\mathcal{C}$\ such that
$\sum_{i=1}^{m}\left\vert H\left(  x_{i}\right)  -F\left(  x_{i}\right)
\right\vert $\ is minimized. \ Then there exists a positive constant $K$ such
that%
\[
\operatorname*{EX}_{x\in\mathcal{D}}\left[  \left\vert H\left(  x\right)
-F\left(  x\right)  \right\vert \right]  \leq\alpha+\inf_{C\in\mathcal{C}%
}\operatorname*{EX}_{x\in\mathcal{D}}\left[  \left\vert C\left(  x\right)
-F\left(  x\right)  \right\vert \right]
\]
with probability at least $1-\delta$ over $X$, provided that%
\[
m\geq\frac{K}{\alpha^{2}}\left(  \operatorname*{fat}\nolimits_{\mathcal{C}%
}\left(  \frac{\alpha}{5}\right)  \log^{2}\frac{1}{\alpha}+\log\frac{1}%
{\delta}\right)  .
\]

\end{theorem}

Theorem \ref{occam2}\ has the following corollary.

\begin{corollary}
\label{occamcor}In the statement of Theorem \ref{occam}, suppose
$\gamma\varepsilon\geq7\eta$. \ Then the bound on $m$ can be replaced by%
\[
m\geq\frac{K}{\gamma^{2}\varepsilon^{2}}\left(  \operatorname*{fat}%
\nolimits_{\mathcal{C}}\left(  \frac{\gamma\varepsilon}{35}\right)  \log
^{2}\frac{1}{\gamma\varepsilon}+\log\frac{1}{\delta}\right)  .
\]

\end{corollary}

Like all proofs in this paper, the proof of Corollary \ref{occamcor} is
deferred to Appendix \ref{PROOFS}.

\subsection{Learning Quantum States\label{LQS}}

We now turn to the problem of learning a quantum state. \ Let $\mathcal{S}$ be
the set of two-outcome measurements on $n$ qubits. \ Also, given an $n$-qubit
mixed state $\rho$, let $F_{\rho}:\mathcal{S}\rightarrow\left[  0,1\right]
$\ be the p-concept defined by $F_{\rho}\left(  E\right)  =\operatorname*{Tr}%
\left(  E\rho\right)  $,\ and let $\mathcal{C}_{n}=\left\{
F_{\rho}\right\} _{\rho}$\ be the class of all such $F_{\rho}$'s. \
Then to apply Theorems \ref{occam} and \ref{occam2}, all we need to
do is upper-bound
$\operatorname*{fat}\nolimits_{\mathcal{C}_{n}}\left(  \gamma\right)
$\ in terms of $n$\ and $\gamma$. \ We will do so using a result of
Ambainis et al.\ \cite{antv}, which upper-bounds the number of
classical bits that can be \textquotedblleft
encoded\textquotedblright\ into $n$ qubits.

\begin{theorem}
[Ambainis et al.\ \cite{antv}]\label{nayakthm}Let $k$ and $n$ be
positive integers with $k>n$. \ For all $k$-bit strings
$y=y_{1}\cdots y_{k}$, let $\rho_{y}$\ be an $n$-qubit mixed state
that \textquotedblleft encodes\textquotedblright\ $y$. \ Suppose
there exist two-outcome measurements $E_{1},\ldots,E_{k}$ such that\
for all $y\in\left\{  0,1\right\}  ^{k}$ and $i\in\left\{
1,\ldots,k\right\}  $,

\begin{enumerate}
\item[(i)] if $y_{i}=0$\ then $\operatorname*{Tr}\left(  E_{i}\rho_{y}\right)
\leq p$, and

\item[(ii)] if $y_{i}=1$\ then $\operatorname*{Tr}\left(  E_{i}\rho
_{y}\right)  \geq1-p$.
\end{enumerate}

Then $n\geq\left(  1-H\left(  p\right)  \right)  k$, where $H$\ is the binary
entropy function.
\end{theorem}

Theorem \ref{nayakthm} has the following easy generalization.

\begin{theorem}
\label{nayakthm2}Let $k$, $n$, and $\left\{  \rho_{y}\right\}  $\ be as in
Theorem \ref{nayakthm}. \ Suppose there exist measurements $E_{1},\ldots
,E_{k}$, as well as real numbers $\alpha_{1},\ldots,\alpha_{k}$, such
that\ for all $y\in\left\{  0,1\right\}  ^{k}$ and $i\in\left\{
1,\ldots,k\right\}  $,

\begin{enumerate}
\item[(i)] if $y_{i}=0$\ then $\operatorname*{Tr}\left(  E_{i}\rho_{y}\right)
\leq\alpha_{i}-\gamma$, and

\item[(ii)] if $y_{i}=1$\ then $\operatorname*{Tr}\left(  E_{i}\rho
_{y}\right)  \geq\alpha_{i}+\gamma$.
\end{enumerate}

Then $n/\gamma^{2}=\Omega\left(  k\right)  $.
\end{theorem}

If we interpret $k$ as the size of a fat-shattered subset of $\mathcal{S}$,
then Theorem \ref{nayakthm2} immediately yields the following upper bound on
fat-shattering dimension.

\begin{corollary}
\label{nayakcor}For all $\gamma>0$ and $n$, we have $\operatorname*{fat}%
\nolimits_{\mathcal{C}_{n}}\left(  \gamma\right)  =O\left(  n/\gamma
^{2}\right)  $.
\end{corollary}

Combining Corollary \ref{occamcor}\ with Corollary \ref{nayakcor}, we find
that if $\gamma\varepsilon\geq7\eta$, then it suffices to use%
\[
m=\left\lceil \frac{K}{\gamma^{2}\varepsilon^{2}}\left(  \operatorname*{fat}%
\nolimits_{\mathcal{C}_{n}}\left(  \frac{\gamma\varepsilon}{35}\right)
\log^{2}\frac{1}{\gamma\varepsilon}+\log\frac{1}{\delta}\right)  \right\rceil
=O\left(  \frac{1}{\gamma^{2}\varepsilon^{2}}\left(  \frac{n}{\gamma
^{2}\varepsilon^{2}}\log^{2}\frac{1}{\gamma\varepsilon}+\log\frac{1}{\delta
}\right)  \right)
\]
measurements. \ Likewise, combining Theorem \ref{occam}\ with Corollary
\ref{nayakcor}, we find that if $\gamma>\eta$, then it suffices to use%
\[
m=\left\lceil \frac{K}{\varepsilon}\left(  \operatorname*{fat}%
\nolimits_{\mathcal{C}_{n}}\left(  \frac{\gamma-\eta}{8}\right)  \log
^{2}\left(  \frac{\operatorname*{fat}\nolimits_{\mathcal{C}_{n}}\left(
\frac{\gamma-\eta}{8}\right)  }{\left(  \gamma-\eta\right)  \varepsilon
}\right)  +\log\frac{1}{\delta}\right)  \right\rceil =O\left(  \frac
{1}{\varepsilon}\left(  \frac{n}{\left(  \gamma-\eta\right)  ^{2}}\log
^{2}\frac{n}{\left(  \gamma-\eta\right)  \varepsilon}+\log\frac{1}{\delta
}\right)  \right)
\]
measurements. \ This completes the proofs of Theorems \ref{qoccam} and
\ref{qoccam2}\ respectively.

\section{Application to Quantum Communication\label{COMM}}

In this section we use our quantum learning theorem to prove a new result
about \textit{one-way communication complexity}. \ Here we consider two
players, Alice and Bob, who hold inputs $x$ and $y$ respectively. \ For
concreteness, let $x$ be an $N$-bit string, and let $y$ be an $M$-bit string.
\ Also, let $f:\mathcal{Z}\rightarrow\left\{  0,1\right\}  $\ be a Boolean
function, where $\mathcal{Z}$ is some subset of $\left\{  0,1\right\}
^{N}\times\left\{  0,1\right\}  ^{M}$.\ \ We call $f$ \textit{total} if
$\mathcal{Z}=\left\{  0,1\right\}  ^{N}\times\left\{  0,1\right\}  ^{M}$, and
\textit{partial} otherwise.

We are interested in the minimum number of bits $k$ that Alice needs to send
to Bob, for Bob to be able to evaluate $f\left(  x,y\right)  $\ for any input
pair $\left(  x,y\right)  \in\mathcal{Z}$. \ We consider three models of
communication: deterministic, randomized, and quantum. \ In the deterministic
model, Alice sends Bob a $k$-bit string $a_{x}$ depending only on $x$. \ Then
Bob, using only $a_{x}$ and $y$, must output $f\left(  x,y\right)  $\ with
certainty. \ In the randomized model, Alice sends Bob a $k$-bit string
$a$\ drawn from a probability distribution $\mathcal{D}_{x}$. \ Then Bob must
output $f\left(  x,y\right)  $\ with probability at least $\frac{2}{3}$\ over
$a\in\mathcal{D}_{x}$.\footnote{We can assume without loss of generality that
Bob is deterministic, i.e. that his output is a function of $a$ and $y$.} \ In
the quantum model, Alice sends Bob a $k$-qubit mixed state $\rho_{x}$. \ Then
Bob, after measuring $\rho_{x}$\ in a basis depending on $y$, must output
$f\left(  x,y\right)  $\ with probability at least $\frac{2}{3}$. \ We use
$\operatorname*{D}\nolimits^{1}\left(  f\right)  $, $\operatorname*{R}%
\nolimits^{1}\left(  f\right)  $, and $\operatorname*{Q}\nolimits^{1}\left(
f\right)  $\ to denote the minimum value of $k$ for which Bob can succeed in
the deterministic, randomized, and quantum models respectively. \ Clearly
$\operatorname*{D}\nolimits^{1}\left(  f\right)  \geq\operatorname*{R}%
\nolimits^{1}\left(  f\right)  \geq\operatorname*{Q}\nolimits^{1}\left(
f\right)  $ for all $f$.

The question that interests us is how small the quantum communication
complexity $\operatorname*{Q}\nolimits^{1}\left(  f\right)  $\ can be compared
to the classical complexities $\operatorname*{D}\nolimits^{1}\left(  f\right)
$\ and $\operatorname*{R}\nolimits^{1}\left(  f\right)  $. \ We know that
there exists a total function $f:\left\{  0,1\right\}  ^{N}\times\left\{
0,1\right\}  ^{N}\rightarrow\left\{  0,1\right\}  $ for which
$\operatorname*{D}\nolimits^{1}\left(  f\right)  =N$\ but $\operatorname*{R}%
\nolimits^{1}\left(  f\right)  =\operatorname*{Q}\nolimits^{1}\left(
f\right)  =O\left(  \log N\right)  $.\footnote{This $f$ is the
equality function: $f\left(  x,y\right)  =1$\ if $x=y$, and $f\left(
x,y\right) =0$\ otherwise.} \ Furthermore, Gavinsky et al.\
\cite{gkkrw} have recently
shown that there exists a partial function $f$ for which $\operatorname*{R}%
\nolimits^{1}\left(  f\right)  =\Omega\left(  \sqrt{N}\right)  $\ but
$\operatorname*{Q}\nolimits^{1}\left(  f\right)  =O\left(  \log N\right)  $.

On the other hand, it follows from a result of Klauck \cite{klauck:cc} that
$\operatorname*{D}\nolimits^{1}\left(  f\right)  =O\left(  M\operatorname*{Q}%
\nolimits^{1}\left(  f\right)  \right)  $ for all total $f$.\ \ Intuitively,
if Bob's input is small, then quantum communication provides at most a limited
savings over classical communication. \ But does the $\operatorname*{D}%
\nolimits^{1}\left(  f\right)  =O\left(  M\operatorname*{Q}\nolimits^{1}%
\left(  f\right)  \right)  $\ bound hold for partial $f$ as well? \ Aaronson
\cite{aar:adv}\ proved a slightly weaker result: for all $f$ (partial or
total), $\operatorname*{D}\nolimits^{1}\left(  f\right)  =O\left(
M\operatorname*{Q}\nolimits^{1}\left(  f\right)  \log\operatorname*{Q}%
\nolimits^{1}\left(  f\right)  \right)  $. \ Whether the $\log
\operatorname*{Q}\nolimits^{1}\left(  f\right)  $\ factor can be removed has
remained an open problem for several years.

Using our quantum learning theorem, we are able to resolve this problem, at
the cost of replacing $\operatorname*{D}\nolimits^{1}\left(  f\right)  $\ by
$\operatorname*{R}\nolimits^{1}\left(  f\right)  $. \ In particular, Theorem
\ref{ccthm}, proved in Appendix \ref{PROOFS}, shows that $\operatorname*{R}%
\nolimits^{1}\left(  f\right)  =O\left(  M\operatorname*{Q}\nolimits^{1}%
\left(  f\right)  \right)  $\ for any Boolean function $f$. \ Also,
Appendix \ref{OPT} uses a recent result of Gavinsky et al.\
\cite{gkkrw}\ to show that Theorem \ref{ccthm} is close to
optimal---and in particular, that it cannot be improved to
$\operatorname*{R}\nolimits^{1}\left(  f\right)  =O\left(
M+\operatorname*{Q}\nolimits^{1}\left(  f\right)  \right)  $.

\section{Open Problems\label{OPEN}}

Perhaps the central question left open by this paper is which classes of
states and measurements can be learned, not only with a linear number of
measurements, but also with a reasonable amount of computation. \ To give two
examples, what is the situation for stabilizer states \cite{ag} or
noninteracting-fermion states \cite{td:fermion}?\footnote{Note that we can
only hope to learn such states efficiently for restricted classes of
measurements. \ Otherwise, even if the state to be learned were a\ classical
basis state $\left\vert x\right\rangle $, a \textquotedblleft
measurement\textquotedblright\ of $\left\vert x\right\rangle $\ might be an
arbitrary polynomial-time computation that fed $x$ as input to a pseudorandom
function.}

On the experimental side, it would be interesting to demonstrate
\textquotedblleft pretty good tomography\textquotedblright\ in photonics, ion
traps, NMR, or any other technology that allows the preparation and
measurement of multi-qubit entangled states. \ Already for three or four
qubits, complete tomography requires hundreds of measurements, and depending
on what accuracy is needed, it seems likely that our learning approach could
yield an efficiency improvement. \ How much of an improvement partly depends
on how far our learning results can be improved, as well as on what the
constant factors are. \ A related issue is that, while one can always reduce
noisy, $k$-outcome measurements to the noiseless, two-outcome measurements
that we consider, one could almost certainly prove better upper bounds by
analyzing realistic measurements more directly.

One might hope for a far-reaching generalization of our learning theorem, to
what is known as \textit{quantum process tomography}. \ Here the goal is to
learn an unknown quantum \textit{operation}\ on $n$ qubits by feeding it
inputs and examining the outputs. \ But for process tomography, it is not hard
to show that exponentially many measurements really are needed; in other
words, the analogue of our learning theorem\ is false.\footnote{Here is a
proof sketch: let $U$ be an $n$-qubit unitary that maps $\left\vert
x\right\rangle \left\vert b\right\rangle $ to $\left\vert x\right\rangle
\left\vert b\oplus f\left(  x\right)  \right\rangle $, for some Boolean
function $f:\left\{  0,1\right\}  ^{n-1}\rightarrow\left\{  0,1\right\}  $.
\ Then to predict $U$ on a $1-\varepsilon$\ fraction of basis states, we need
to know $\left(  1-\varepsilon\right)  2^{n-1}$\ bits of the truth table of
$f$. \ But Holevo's Theorem implies that, by examining $U\left\vert \psi
_{i}\right\rangle $ for $T$ input states $\left\vert \psi_{1}\right\rangle
,\ldots,\left\vert \psi_{T}\right\rangle $, we can learn at most $Tn$\ bits
about $f$.} \ Still, it would be interesting to know if there is anything to
say about \textquotedblleft pretty good process tomography\textquotedblright%
\ for restricted classes of operations.

Finally, our quantum information results immediately suggest several problems.
\ First, does $\mathsf{BQP/qpoly}=\mathsf{YQP/poly}$? \ In other words, can we
use classical advice to verify quantum advice even in the worst-case setting?
\ Alternatively, can we give a \textquotedblleft quantum
oracle\textquotedblright\ (see \cite{ak}) relative to which
$\mathsf{BQP/qpoly}\neq\mathsf{YQP/poly}$? \ Second, can the relation
$\operatorname*{R}\nolimits^{1}\left(  f\right)  =O\left(  M\operatorname*{Q}%
\nolimits^{1}\left(  f\right)  \right)  $\ be improved to $\operatorname*{D}%
^{1}\left(  f\right)  =O\left(  M\operatorname*{Q}\nolimits^{1}\left(
f\right)  \right)  $\ for all $f$? \ Perhaps learning theory techniques could
even shed light on the old problem of whether $\operatorname*{R}%
\nolimits^{1}\left(  f\right)  =O\left(  \operatorname*{Q}\nolimits^{1}\left(
f\right)  \right)  $\ for all total $f$.

\section{Acknowledgments}

I thank Noga Alon, Dave Bacon, Peter Bartlett, Robin Blume-Kohout, Andy
Drucker, Aram Harrow, Tony Leggett, Peter Shor, Luca Trevisan, Umesh Vazirani,
Ronald de Wolf, and the anonymous reviewers of an earlier draft for helpful
discussions and correspondence.

\bibliographystyle{plain}
\bibliography{thesis}

\section{\label{OPT}Appendix: Optimality of Theorem \ref{ccthm}}

It is easy to see that, in Theorem \ref{ccthm}, the upper bound on
$\operatorname*{R}\nolimits^{1}\left(  f\right)  $\ needs to depend both on
$M$ and on $\operatorname*{Q}\nolimits^{1}\left(  f\right)  $. \ For the index
function\footnote{This is the function $f:\left\{  0,1\right\}  ^{N}%
\times\left\{  1,\ldots,N\right\}  \rightarrow\left\{  0,1\right\}  $\ defined
by $f\left(  x_{1}\cdots x_{N},i\right)  =x_{i}$.} yields a total $f$ for
which $\operatorname*{R}\nolimits^{1}\left(  f\right)  $\ is exponentially
larger than $M$, while the recent results of Gavinsky et al.\ \cite{gkkrw}%
\ yield a partial $f$ for which $\operatorname*{R}\nolimits^{1}\left(
f\right)  $\ is exponentially larger than $\operatorname*{Q}\nolimits^{1}%
\left(  f\right)  $. \ However, is it possible that Theorem \ref{ccthm} could
be improved to $\operatorname*{R}\nolimits^{1}\left(  f\right)  =O\left(
M+\operatorname*{Q}\nolimits^{1}\left(  f\right)  \right)  $?

Using a slight generalization of Gavinsky et al.'s result, we are able to rule
out this possibility. \ Gavinsky et al.\ consider the following one-way
communication problem, called the \textit{Boolean Hidden Matching Problem}.
\ Alice is given a string $x\in\left\{  0,1\right\}  ^{N}$. \ For some
parameter $\alpha>0$, Bob is given $\alpha N$\ disjoint edges $\left(
i_{1},j_{1}\right)  ,\ldots,\left(  i_{\alpha N},j_{\alpha N}\right)  $\ in
$\left\{  1,\ldots,N\right\}  ^{2}$, together with a string $w\in\left\{
0,1\right\}  ^{\alpha N}$. \ (Thus Bob's input length is $M=O\left(  \alpha
N\log N\right)  $.) \ Alice and Bob are promised that either

\begin{enumerate}
\item[(i)] $x_{i_{\ell}}\oplus x_{j_{\ell}}\equiv w_{\ell}\left(
\operatorname{mod}2\right)  $\ for all $\ell\in\left\{  1,\ldots,\alpha
N\right\}  $, or

\item[(ii)] $x_{i_{\ell}}\oplus x_{j_{\ell}}\not \equiv w_{\ell}\left(
\operatorname{mod}2\right)  $\ for all $\ell\in\left\{  1,\ldots,\alpha
N\right\}  $.
\end{enumerate}

Bob's goal is to output $f=0$\ in case (i), or $f=1$\ in case (ii).

It is not hard to see that $\operatorname*{Q}\nolimits^{1}\left(  f\right)
=O\left(  \frac{1}{\alpha}\log N\right)  $\ for all $\alpha>0$.\footnote{The
protocol is as follows: first Alice sends the $\log N$-qubit quantum message
$\frac{1}{\sqrt{N}}\sum_{i=1}^{N}\left(  -1\right)  ^{x_{i}}\left\vert
i\right\rangle $. \ Then Bob measures in a basis corresponding to $\left(
i_{1},j_{1}\right)  ,\ldots,\left(  i_{\alpha N},j_{\alpha N}\right)  $.
\ With probability $2\alpha$, Bob will learn whether $x_{i_{\ell}}\oplus
x_{j_{\ell}}\equiv w_{\ell}$\ for some edge $\left(  i_{\ell},j_{\ell}\right)
$. \ So it suffices to amplify the protocol $O\left(  1/\alpha\right)
$\ times.} \ What Gavinsky et al.\ showed is that, if $\alpha\approx
1/\sqrt{\log N}$, then $\operatorname*{R}\nolimits^{1}\left(  f\right)
=\Omega\left(  \sqrt{N/\alpha}\right)  $. \ By tweaking their proof a bit, one
can generalize their result to $\operatorname*{R}\nolimits^{1}\left(
f\right)  =\Omega\left(  \sqrt{N/\alpha}\right)  $\ for \textit{all}
$\alpha\ll1/\sqrt{\log N}$.\footnote{R. de Wolf, personal communication.} \ So
in particular, set\ $\alpha:=1/\sqrt{N}$. \ Then we obtain a partial Boolean
function $f$ for which $M=O\left(  \sqrt{N}\log N\right)  $\ and
$\operatorname*{Q}\nolimits^{1}\left(  f\right)  =O\left(  \sqrt{N}\log
N\right)  $\ but $\operatorname*{R}\nolimits^{1}\left(  f\right)
=\Omega\left(  N^{3/4}\right)  $, thereby refuting the conjecture that
$\operatorname*{R}\nolimits^{1}\left(  f\right)  =O\left(  M+\operatorname*{Q}%
\nolimits^{1}\left(  f\right)  \right)  $.

As a final remark, the Boolean Hidden Matching Problem clearly satisfies
$\operatorname*{D}\nolimits^{1}\left(  f\right)  =\Omega\left(  N\right)
$\ for all $\alpha>0$. \ So by varying $\alpha$, we immediately get not only
that $\operatorname*{D}\nolimits^{1}\left(  f\right)  =O\left(
M+\operatorname*{Q}\nolimits^{1}\left(  f\right)  \right)  $\ is false, but
that Aaronson's bound $\operatorname*{D}\nolimits^{1}\left(  f\right)
=O\left(  M\operatorname*{Q}\nolimits^{1}\left(  f\right)  \log
\operatorname*{Q}\nolimits^{1}\left(  f\right)  \right)  $\ \cite{aar:adv}\ is
\textit{tight} up to a polylogarithmic term. \ This answers one of the open
questions in \cite{aar:adv}.

\section{Appendix: Application to Quantum Advice\label{ADV}}

Having applied our quantum learning theorem to communication complexity, in
this appendix we apply the theorem to computational complexity. \ In
particular, we will show how to use a trusted classical string to perform
approximate verification of an untrusted quantum state.

The following conventions will be helpful throughout the section. \ We
identify a language $L\subseteq\left\{  0,1\right\}  ^{\ast}$ with the Boolean
function $L:\left\{  0,1\right\}  ^{\ast}\rightarrow\left\{  0,1\right\}
$\ such that $L\left(  x\right)  =1$\ if and only if $x\in L$. \ Given a
quantum algorithm $A$, we let $P_{A}^{1}\left(  \left\vert \psi\right\rangle
\right)  $\ be the probability that $A$\ accepts and $P_{A}^{0}\left(
\left\vert \psi\right\rangle \right)  $\ be the probability that $A$\ rejects
if given the state $\left\vert \psi\right\rangle $\ as input. \ Note that $A$
might neither accept nor reject (in other words, output \textquotedblleft
don't know\textquotedblright), in which case $P_{A}^{0}\left(  \left\vert
\psi\right\rangle \right)  +P_{A}^{1}\left(  \left\vert \psi\right\rangle
\right)  <1$. \ Finally, we use $\mathcal{H}_{2}^{\otimes k}$\ to denote a
Hilbert space of $k$ qubits,\ and $\operatorname*{poly}\left(  n\right)  $\ to
denote an arbitrary polynomial in $n$.

\subsection{Quantum Advice and Proofs\label{QAP}}

Recall that $\mathsf{BQP}$, or Bounded-Error Quantum Polynomial-Time, is the
class of problems efficiently solvable by a quantum computer. \ Then
$\mathsf{BQP/qpoly}$\ is a generalization of $\mathsf{BQP}$, in which the
quantum computer is given a\ polynomial-size \textquotedblleft quantum advice
state\textquotedblright\ that depends only on the input length $n$, but could
otherwise be arbitrarily hard to prepare. \ More formally:

\begin{definition}
A language $L\subseteq\left\{  0,1\right\}  ^{\ast}$\ is
in\ $\mathsf{BQP/qpoly}$\ if there exists a polynomial-time quantum algorithm
$A$ such that for all input lengths $n$, there exists a quantum advice state
$\left\vert \psi_{n}\right\rangle \in\mathcal{H}_{2}^{\otimes
\operatorname*{poly}\left(  n\right)  }$\ such that $P_{A}^{L\left(  x\right)
}\left(  \left\vert x\right\rangle \left\vert \psi_{n}\right\rangle \right)
\geq\frac{2}{3}$\ for all $x\in\left\{  0,1\right\}  ^{n}$.
\end{definition}

How powerful is this class? \ Aaronson \cite{aar:adv}\ proved the first
limitation on $\mathsf{BQP/qpoly}$,\ by showing that $\mathsf{BQP/qpoly}%
\subseteq\mathsf{PostBQP/poly}$. \ Here $\mathsf{PostBQP}$\ is a
generalization of $\mathsf{BQP}$\ in which we can \textquotedblleft
postselect\textquotedblright\ on the outcomes of measurements,\footnote{See
\cite{aar:pp} for a detailed definition, as well as a proof that
$\mathsf{PostBQP}$ coincides with the classical complexity class $\mathsf{PP}%
$.} and $\mathsf{/poly}$\ means \textquotedblleft with polynomial-size
classical advice.\textquotedblright\ \ Intuitively, this result means that
anything we can do with quantum advice, we can also do with classical advice,
provided we are willing to use exponentially more computation time to extract
what the advice is telling us.

In addition to quantum advice, we will also be interested in quantum proofs.
\ Compared to advice, a proof has the advantage that it can be tailored to a
particular input $x$, but\ the disadvantage that it cannot be trusted. \ In
other words, while an advisor's only goal is to help the algorithm $A$ decide
whether $x\in L$, a prover wants to \textit{convince} $A$ that $x\in L$. \ The
class of problems that admit polynomial-size quantum proofs is called
$\mathsf{QMA}$\ (Quantum Merlin-Arthur).

\begin{definition}
A language $L$\ is in\ $\mathsf{QMA}$ if there exists a polynomial-time
quantum algorithm $A$ such that for all $x\in\left\{  0,1\right\}  ^{n}$:

\begin{enumerate}
\item[(i)] If $x\in L$\ then there exists a quantum witness $\left\vert
\varphi\right\rangle \in\mathcal{H}_{2}^{\otimes\operatorname*{poly}\left(
n\right)  }$\ such that $P_{A}^{1}\left(  \left\vert x\right\rangle \left\vert
\varphi\right\rangle \right)  \geq\frac{2}{3}$.

\item[(ii)] If $x\notin L$\ then $P_{A}^{1}\left(  \left\vert x\right\rangle
\left\vert \varphi\right\rangle \right)  \leq\frac{1}{3}$\ for all $\left\vert
\varphi\right\rangle $.
\end{enumerate}
\end{definition}

One can think of $\mathsf{QMA}$ as a quantum analogue of $\mathsf{NP}$.

\subsection{Untrusted Advice\label{YP}}

To state our result in the strongest possible way, we need to define a new
notion called \textit{untrusted advice}, which might be of independent
interest for complexity theory. \ Intuitively, untrusted advice is a
\textquotedblleft hybrid\textquotedblright\ of proof and advice: it is like a
proof in that it cannot be trusted, but like advice in that depends only on
the input length $n$. \ More concretely, let us define the complexity class
$\mathsf{YP}$, or \textquotedblleft Yoda Polynomial-Time,\textquotedblright%
\ to consist of all problems solvable in classical polynomial time with help
from polynomial-size untrusted advice:\footnote{Here Yoda, from \textit{Star
Wars}, is intended to evoke a sage whose messages are highly generic
(\textquotedblleft Do or do not... there is no try\textquotedblright). \ One
motivation for the name $\mathsf{YP}$\ is that, to our knowledge, there had
previously been no complexity class starting with a `Y'.}

\begin{definition}
\label{ypdef}A language $L$\ is in\ $\mathsf{YP}$\ if there exists a
polynomial-time algorithm $A$ such that for all $n$:

\begin{enumerate}
\item[(i)] There exists a string $y_{n}\in\left\{  0,1\right\}  ^{p\left(
n\right)  }$ such that $A\left(  x,y_{n}\right)  $\ outputs $L\left(
x\right)  $\ for all $x\in\left\{  0,1\right\}  ^{n}$.

\item[(ii)] $A\left(  x,y\right)  $ outputs either $L\left(  x\right)  $\ or
\textquotedblleft don't know\textquotedblright\ for all $x\in\left\{
0,1\right\}  ^{n}$\ and all $y$.
\end{enumerate}
\end{definition}

From the definition, it is clear that $\mathsf{YP}$\ is contained both in
$\mathsf{P/poly}$\ and in $\mathsf{NP\cap coNP}$. \ Indeed, while we are at
it, let us initiate the study of $\mathsf{YP}$, by mentioning four simple
facts that relate $\mathsf{YP}$\ to standard complexity classes.

\begin{theorem}
\label{ypthm}\quad

\begin{enumerate}
\item[(i)] $\mathsf{ZPP}\subseteq\mathsf{YP}$.

\item[(ii)] $\mathsf{YE}=\mathsf{NE}\cap\mathsf{coNE}$, where $\mathsf{YE}%
$\ is the exponential-time analogue of $\mathsf{YP}$\ (i.e., both the advice
size and the verifier's running time are $2^{O\left(  n\right)  }$).

\item[(iii)] If $\mathsf{P}=\mathsf{YP}$\ then $\mathsf{E}=\mathsf{NE}%
\cap\mathsf{coNE}$.

\item[(iv)] If $\mathsf{E}=\mathsf{NE}^{\mathsf{NP}^{\mathsf{NP}}}$\ then
$\mathsf{P}=\mathsf{YP}$.
\end{enumerate}
\end{theorem}

Naturally one can also define $\mathsf{YPP}$\ and $\mathsf{YQP}$, the
(bounded-error) probabilistic and quantum analogues of $\mathsf{YP}$. \ For
brevity, we give only the definition of $\mathsf{YQP}$.

\begin{definition}
\label{yqpdef}A language $L$\ is in\ $\mathsf{YQP}$\ if there exists a
polynomial-time quantum algorithm $A$ such that for all $n$:

\begin{enumerate}
\item[(i)] There exists a state $\left\vert \varphi_{n}\right\rangle
\in\mathcal{H}_{2}^{\otimes\operatorname*{poly}\left(  n\right)  }$ such that
$P_{A}^{L\left(  x\right)  }\left(  \left\vert x\right\rangle \left\vert
\varphi_{n}\right\rangle \right)  \geq\frac{2}{3}$\ for all $x\in\left\{
0,1\right\}  ^{n}$.

\item[(ii)] $P_{A}^{1-L\left(  x\right)  }\left(  \left\vert x\right\rangle
\left\vert \varphi\right\rangle \right)  \leq\frac{1}{3}$ for all
$x\in\left\{  0,1\right\}  ^{n}$ and all $\left\vert \varphi\right\rangle $.
\end{enumerate}
\end{definition}

\begin{figure}
[ptb]
\begin{center}
\includegraphics[
trim=0.538979in 1.758652in 0.542857in 3.111541in,
height=1.9801in,
width=2.406in
]%
{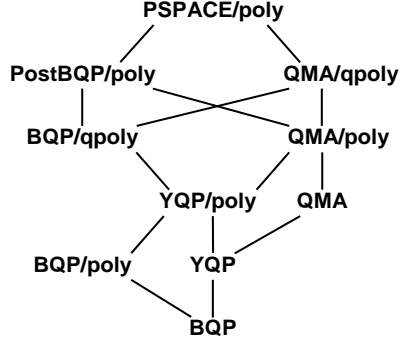}%
\caption{Some quantum advice and proof classes. \ The containment
$\mathsf{BQP/qpoly}\subseteq\mathsf{PostBQP/poly}$\ was shown in
\cite{aar:adv}, while $\mathsf{QMA/qpoly}\subseteq\mathsf{PSPACE/poly}$\ was
shown in \cite{aar:qmaqpoly}.}%
\label{incl}%
\end{center}
\end{figure}

By analogy to the classical case, $\mathsf{YQP}$ is contained both in
$\mathsf{BQP/qpoly}$\ and in$\ \mathsf{QMA\cap coQMA}$. \ We also have
$\mathsf{YQP/qpoly}=\mathsf{BQP/qpoly}$, since the untrusted $\mathsf{YQP}%
$\ advice can be tacked onto the trusted $\mathsf{/qpoly}$ advice. \ Figure
\ref{incl} shows the known containments among various classes involving
quantum advice and proofs.%

\subsection{Heuristic Complexity\label{HEUR}}

Ideally, we would like to show that $\mathsf{BQP/qpoly}=\mathsf{YQP/poly}%
$---in other words, that trusted quantum advice can be replaced by trusted
classical advice together with untrusted quantum advice. \ However, we will
only be able to prove this for the \textit{heuristic} versions of these
classes: that is, the versions where we allow algorithms that can err on some
fraction of inputs.\footnote{Closely related to heuristic complexity is the
better-known \textit{average-case} complexity. \ In average-case complexity
one considers algorithms that can never err, but that are allowed to output
\textquotedblleft don't know\textquotedblright\ on some fraction of inputs.}
\ We now explain what this means (for details, see the excellent survey by
Bogdanov and Trevisan \cite{bt}).

A \textit{distributional problem} is a pair $\left(  L,\left\{  \mathcal{D}%
_{n}\right\}  \right)  $, where $L\subseteq\left\{  0,1\right\}  ^{\ast}$\ is
a language and $\mathcal{D}_{n}$\ is a probability distribution over $\left\{
0,1\right\}  ^{n}$. \ Intuitively, for each input length $n$, the goal will be
to decide whether $x\in L$\ with high probability over $x$ drawn from
$\mathcal{D}_{n}$. \ In particular, the class $\mathsf{HeurP}$, or
Heuristic-$\mathsf{P}$, consists (roughly speaking) of all distributional
problems that can be solved in polynomial time on a $1-\frac{1}%
{\operatorname*{poly}\left(  n\right)  }$\ fraction of inputs.

\begin{definition}
A distributional problem $\left(  L,\left\{  \mathcal{D}_{n}\right\}  \right)
$\ is in $\mathsf{HeurP}$\ if there exists a polynomial-time algorithm $A$
such that for all $n$ and $\varepsilon>0$:%
\[
\Pr_{x\in\mathcal{D}_{n}}\left[  A\left(  x,0^{\left\lceil 1/\varepsilon
\right\rceil }\right)  \text{ outputs }L\left(  x\right)  \right]
\geq1-\varepsilon.
\]

\end{definition}

One can also define $\mathsf{HeurP/poly}$, or $\mathsf{HeurP}$\ with
polynomial-size advice. \ (Note that in this context, \textquotedblleft
polynomial-size\textquotedblright\ means polynomial not just in $n$ but in
$1/\varepsilon$ as well.) \ Finally, let us define the heuristic analogues of
$\mathsf{BQP}$\ and $\mathsf{YQP}$.

\begin{definition}
A distributional problem $\left(  L,\left\{  \mathcal{D}_{n}\right\}  \right)
$\ is in $\mathsf{HeurBQP}$\ if there exists a polynomial-time quantum
algorithm $A$ such that for all $n$ and $\varepsilon>0$:%
\[
\Pr_{x\in\mathcal{D}_{n}}\left[  P_{A}^{L\left(  x\right)  }\left(  \left\vert
x\right\rangle \left\vert 0\right\rangle ^{\otimes\left\lceil 1/\varepsilon
\right\rceil }\right)  \geq\frac{2}{3}\right]  \geq1-\varepsilon.
\]

\end{definition}

\begin{definition}
A distributional problem $\left(  L,\left\{  \mathcal{D}_{n}\right\}  \right)
$\ is in $\mathsf{HeurYQP}$\ if there exists a polynomial-time quantum
algorithm $A$ such that for all $n$ and $\varepsilon>0$:

\begin{enumerate}
\item[(i)] There exists a state $\left\vert \varphi_{n,\varepsilon
}\right\rangle \in\mathcal{H}_{2}^{\otimes\operatorname*{poly}\left(
n,1/\varepsilon\right)  }$\ such that%
\[
\Pr_{x\in\mathcal{D}_{n}}\left[  P_{A}^{L\left(  x\right)  }\left(  \left\vert
x\right\rangle \left\vert \varphi_{n,\varepsilon}\right\rangle \right)
\geq\frac{2}{3}\right]  \geq1-\varepsilon.
\]

\item[(ii)] The probability over $x\in\mathcal{D}_{n}$\ that there exists a
$\left\vert \varphi\right\rangle $\ such that $P_{A}^{1-L\left(  x\right)
}\left(  \left\vert x\right\rangle \left\vert \varphi\right\rangle \right)
\geq\frac{1}{3}$\ is at most $\varepsilon$.
\end{enumerate}
\end{definition}

It is clear that$\ \mathsf{HeurYQP/poly}\subseteq\mathsf{HeurBQP/qpoly}%
=\mathsf{HeurYQP/qpoly}$.

\subsection{Overview of Proof\label{WITNESS}}

Our goal is to show that $\mathsf{HeurBQP/qpoly}=\mathsf{HeurYQP/poly}$: in
the heuristic setting, trusted classical advice can be used to verify
untrusted quantum advice. \ The intuition behind this result is simple: the
classical advice to the $\mathsf{HeurYQP}$\ verifier $V$ will consist of a
polynomial number of randomly-chosen \textquotedblleft test
inputs\textquotedblright\ $x_{1},\ldots,x_{m}$, as well as whether each
$x_{i}$\ belongs to the language $L$. \ Then given an untrusted quantum advice
state $\left\vert \varphi\right\rangle $, first $V$ will check that
$\left\vert \varphi\right\rangle $\ yields the correct answers on
$x_{1},\ldots,x_{m}$; only if $\left\vert \varphi\right\rangle $\ passes this
initial test will $V$ use it on the input $x$ of interest. \ By appealing to
our quantum learning theorem, we will argue that any $\left\vert
\varphi\right\rangle $\ that passes the initial test must yield the correct
answers for \textit{most} $x$ with high probability.

But there is a problem: what if a dishonest prover sends a state $\left\vert
\varphi\right\rangle $\ such that, while $V$'s measurements succeed in
\textquotedblleft verifying\textquotedblright\ $\left\vert \varphi
\right\rangle $, they also \textit{corrupt} it? \ Indeed, even if $V$ repeats
the verification procedure many times, conceivably $\left\vert \varphi
\right\rangle $\ could be corrupted by the very last repetition without $V$
ever realizing it. \ Intuitively, the easiest way to avoid this problem is
just to repeat the verification procedure a random number of times. \ To
formalize this intuition, we need the following \textquotedblleft quantum
union bound,\textquotedblright\ which was proved by Aaronson
\cite{aar:qmaqpoly} based on a result of Ambainis et al.\ \cite{antv}.

\begin{proposition}
[Aaronson \cite{aar:qmaqpoly}]\label{qunion}Let $E_{1},\ldots,E_{m}$\ be
two-outcome measurements,\ and suppose $\operatorname*{Tr}\left(  E_{i}%
\rho\right)  \geq1-\epsilon$\ for all $i\in\left\{  1,\ldots,m\right\}  $.
\ Then if we apply $E_{1},\ldots,E_{m}$\ in sequence to the initial state
$\rho$, the probability that any of the $E_{i}$'s reject is at most
$m\sqrt{\epsilon}$.
\end{proposition}

Using Proposition \ref{qunion},\ we can prove the following \textquotedblleft
Witness Protection Lemma.\textquotedblright

\begin{lemma}
[Witness Protection Lemma]\label{thelem}Let $\mathcal{E}=\left\{  E_{1}%
,\ldots,E_{m}\right\}  $\ be a set of two-outcome measurements,\ and let $T$
be a positive integer. \ Then there exists a test procedure $Q$ with the
following properties:

\begin{enumerate}
\item[(i)] $Q$ takes a state $\rho_{0}$\ as input, applies at most
$T$\ measurements from $\mathcal{E}$, and then returns either
\textquotedblleft success\textquotedblright\ or \textquotedblleft
failure.\textquotedblright

\item[(ii)] If $\operatorname*{Tr}\left(  E_{i}\rho_{0}\right)  \geq
1-\epsilon$\ for all $i$, then $Q$ succeeds\ with probability at least
$1-T\sqrt{\epsilon}$.

\item[(iii)] If $Q$ succeeds\ with probability at least $\lambda$, then
conditioned on succeeding, $Q$ outputs a state $\sigma$\ such that
$\operatorname*{Tr}\left(  E_{i}\sigma\right)  \geq1-2\sqrt{\frac{m}{\lambda
T}}$\ for all $i$.
\end{enumerate}
\end{lemma}

Finally, by using Lemma \ref{thelem}, we can prove Theorem \ref{advthm}: that
$\mathsf{HeurBQP/qpoly}=\mathsf{HeurYQP/poly}\subseteq\mathsf{HeurQMA/poly}$.

\section{Appendix: Learning from Measurement Results\label{MEAS}}

In Section \ref{SC} we considered a model where for each measurement $E$, the
learner is told the approximate value of $\operatorname*{Tr}\left(
E\rho\right)  $. \ This model suffices for our applications to quantum
computing. \ But for other applications, it might be natural to ask what
happens if we instead assume that for each $E$, the learner is merely given
a\textit{ }measurement outcome: that is, a bit that is $1$ with probability
$\operatorname*{Tr}\left(  E\rho\right)  $\ and $0$ with probability
$1-\operatorname*{Tr}\left(  E\rho\right)  $. \ Of course, if the learner were
given many such measurement outcomes for the same $E$, it could form an
estimate of $\operatorname*{Tr}\left(  E\rho\right)  $. \ But we are assuming
that for each $E$, the learner only receives one measurement outcome.

We will show that, even in this seemingly weak model, an $n$-qubit quantum
state can still be learned using $O\left(  n\right)  $ measurements, although
the dependence on the parameters $\gamma$\ and $\varepsilon$\ will worsen.

The general task we are considering---that of learning a p-concept given only
samples from its associated probability distribution---is called
\textit{learning in the p-concept model}. \ The task was first studied in the
early 1990's by Kearns and Schapire \cite{ks:pconcept}, who left open whether
it can always be done if the fat-shattering dimension is finite. \ Alon et
al.\ \cite{abch} answered this question affirmatively in a breakthrough a few
years later. \ Unfortunately, Alon et al.\ never worked out the\ actual
complexity bound implied by their result, and to the best of our knowledge no
one else did either. \ Thus, our first task will be to fill this rather large
gap in the literature. \ We will do so using Theorem \ref{occam2}, which was
proven by Bartlett and Long \cite{bartlettlong}\ building on ideas of Alon et al.

\begin{theorem}
\label{alonthm}Let $\mathcal{S}$\ be a sample space, let $\mathcal{C}$\ be a
p-concept class over $\mathcal{S}$,\ and let $\mathcal{D}$\ be a probability
measure over $\mathcal{S}$. \ Fix a p-concept $F\in\mathcal{C}$, as well as
error parameters $\varepsilon,\gamma>0$.\ \ Suppose we are given $m$
samples\ $X=\left(  x_{1},\ldots,x_{m}\right)  $\ drawn independently from
$\mathcal{D}$, as well as bits $B=\left(  b_{1},\ldots,b_{m}\right)  $\ such
that each $b_{i}$\ is $1$ with independent probability $F\left(  x_{i}\right)
$. \ Suppose also that we choose a hypothesis $H\in\mathcal{C}$\ to\ minimize
the quadratic functional $\sum_{i=1}^{m}\left(  H\left(  x_{i}\right)
-b_{i}\right)  ^{2}$. \ Then there exists a positive constant $K$ such that%
\[
\Pr_{x\in\mathcal{D}}\left[  \left\vert H\left(  x\right)  -F\left(  x\right)
\right\vert >\gamma\right]  \leq\varepsilon
\]
with probability at least $1-\delta$\ over $X$ and $B$, provided that%
\[
m\geq\frac{K}{\gamma^{4}\varepsilon^{2}}\left(  \operatorname*{fat}%
\nolimits_{\mathcal{C}}\left(  \frac{\gamma^{2}\varepsilon}{10}\right)
\log^{2}\frac{1}{\gamma\varepsilon}+\log\frac{1}{\delta}\right)  .
\]

\end{theorem}

We can now prove Theorem \ref{qoccam3}: that%
\[
m=O\left(  \frac{1}{\gamma^{4}\varepsilon^{2}}\left(  \frac{n}{\gamma
^{4}\varepsilon^{2}}\log^{2}\frac{1}{\gamma\varepsilon}+\log\frac{1}{\delta
}\right)  \right)
\]
measurements suffice to learn an $n$-qubit quantum state in the p-concept
model. \ As in Section \ref{SC}, let $\mathcal{C}_{n}$\ be the class of
functions $f_{\rho}:\mathcal{S}\rightarrow\left[  0,1\right]  $\ defined by
$f_{\rho}\left(  E\right)  =\operatorname*{Tr}\left(  E\rho\right)  $. \ Then
the theorem follows immediately from Theorem \ref{alonthm}, together with the
fact that%
\[
\operatorname*{fat}\nolimits_{\mathcal{C}_{n}}\left(  \frac{\gamma
^{2}\varepsilon}{10}\right)  =O\left(  \frac{n}{\left(  \gamma^{2}%
\varepsilon/10\right)  ^{2}}\right)  =O\left(  \frac{n}{\gamma^{4}%
\varepsilon^{2}}\right)
\]
by Corollary \ref{nayakcor}.

Given a state $\rho$, Theorem \ref{qoccam3}\ upper-bounds the number of
measurements needed to estimate the measurement probabilities
$\operatorname*{Tr}\left(  E\rho\right)  $. \ Can we do better if, instead of
estimating the probabilities, we merely want to\ \textit{predict} the outcomes
themselves with nontrivial bias? \ In Appendix \ref{PRED}, we will prove an
almost-tight variant of Theorem \ref{qoccam3}\ that is optimized for this
prediction task.

\section{Appendix: Lower Bounds\label{LB}}

Having proved upper bounds on the measurement complexity of quantum learning,
in this appendix we turn to lower bounds. \ Roughly speaking, we will show
that there exists a measurement distribution $\mathcal{D}$ for which%
\[
m=\Omega\left(  \frac{1}{\varepsilon}\left(  \frac{n}{\gamma^{2}}+\log\frac
{1}{\delta}\right)  \right)
\]
measurements are necessary to learn an $n$-qubit state. \ Also, in the model
of Appendix \ref{MEAS}---the model where each measurement is applied only
once---we will show that%
\[
m=\Omega\left(  \frac{1}{\varepsilon}\left(  \frac{n}{\gamma^{4}}+\log\frac
{1}{\delta}\right)  \right)
\]
measurements are necessary.\footnote{Anthony and Bartlett \cite{ab:learn}%
\ proved a generic lower bound on sample complexity in terms of fat-shattering
dimension. \ However, their bound only implies that%
\[
m=\Omega\left(  \frac{1}{\varepsilon}\left(  \frac{n/\gamma^{2}}{\log
^{2}\left(  n/\gamma^{2}\right)  }+\log\frac{1}{\delta}\right)  \right)
\]
measurements are necessary. \ Using an argument more tailored to our problem,
we were able to get rid of the $\log^{2}\left(  n/\gamma^{2}\right)
$\ factor, improving the dependence on $\operatorname*{fat}_{\mathcal{C}_{n}%
}\left(  \gamma\right)  \sim n/\gamma^{2}$\ to linear.} \ In particular, this
means that Theorem \ref{qoccam} is tight in its dependence on $n$, while
Theorem \ref{qoccam2}\ is tight up to a multiplicative factor of $\log
^{2}\left(  n/\gamma\right)  $.

More formally:

\begin{theorem}
\label{lower}Fix an integer $n>0$ and error parameters $\varepsilon
,\delta,\gamma\in\left(  0,1\right)  $. \ Then there exists a distribution
$\mathcal{D}$\ over $n$-qubit measurements for which the following holds.

\begin{enumerate}
\item[(i)] Suppose%
\[
m=o\left(  \frac{1}{\varepsilon}\left(  \frac{n}{\gamma^{2}}+\log\frac
{1}{\delta}\right)  \right)  .
\]
Then there is no learning algorithm that, given measurements $\mathcal{E}%
=\left(  E_{1},\ldots,E_{m}\right)  $\ drawn independently from $\mathcal{D}$,
as well as real numbers $p_{1},\ldots,p_{m}$\ such that $\left\vert
p_{i}-\operatorname*{Tr}\left(  E_{i}\rho\right)  \right\vert \leq\gamma
^{2}/n$ for all $i$, outputs a hypothesis state $\sigma$\ such that%
\[
\Pr_{E\in\mathcal{D}}\left[  \left\vert \operatorname*{Tr}\left(
E\sigma\right)  -\operatorname*{Tr}\left(  E\rho\right)  \right\vert
>\gamma\right]  \leq\varepsilon
\]
with probability at least $1-\delta$\ over $\mathcal{E}$.

\item[(ii)] Suppose%
\[
m=o\left(  \frac{1}{\varepsilon}\left(  \frac{n}{\gamma^{4}}+\log\frac
{1}{\delta}\right)  \right)  .
\]
Then there is no learning algorithm that, given measurements $\mathcal{E}%
=\left(  E_{1},\ldots,E_{m}\right)  $\ drawn independently from $\mathcal{D}$,
as well as bits $B=\left(  b_{1},\ldots,b_{m}\right)  $ where each $b_{i}$\ is
$1$ with independent probability $\operatorname*{Tr}\left(  E_{i}\rho\right)
$, outputs a hypothesis state $\sigma$\ such that%
\[
\Pr_{E\in\mathcal{D}}\left[  \left\vert \operatorname*{Tr}\left(
E\sigma\right)  -\operatorname*{Tr}\left(  E\rho\right)  \right\vert
>\gamma\right]  \leq\varepsilon
\]
with probability at least $1-\delta$\ over $\mathcal{E}$ and $B$.
\end{enumerate}
\end{theorem}

To prove Theorem \ref{lower}, it will be helpful to introduce a new parameter
that we call the \textit{fine-shattering dimension}. \ This parameter is like
the fat-shattering dimension but with additional restrictions.

\begin{definition}
\label{fine}Let $\mathcal{S}$\ be a sample space, let $\mathcal{C}$ be a
p-concept class over $\mathcal{S}$, and let $0<\gamma\leq\frac{1}{2}$\ and
$\eta\geq0$ be real numbers. \ We say a set $\left\{  s_{1},\ldots
,s_{k}\right\}  \subseteq\mathcal{S}$\ is $\left(  \gamma,\eta\right)
$-fine-\textit{shattered} by $\mathcal{C}$ if for all $B\subseteq\left\{
1,\ldots,k\right\}  $, there exists a p-concept $F\in\mathcal{C}$\ such that
for all $i\in\left\{  1,\ldots,k\right\}  $,

\begin{enumerate}
\item[(i)] if $i\notin B$ then $\frac{1}{2}-\gamma-\eta\leq F\left(
s_{i}\right)  \leq\frac{1}{2}-\gamma$, and

\item[(ii)] if $i\in B$ then $\frac{1}{2}+\gamma\leq F\left(  s_{i}\right)
\leq\frac{1}{2}+\gamma+\eta$.
\end{enumerate}

Then the $\left(  \gamma,\eta\right)  $-fine-\textit{shattering dimension} of
$\mathcal{C}$, or $\operatorname*{fine}_{\mathcal{C}}\left(  \gamma
,\eta\right)  $, is the maximum $k$ such that some subset $\left\{
s_{1},\ldots,s_{k}\right\}  $ of $\mathcal{S}$ is $\left(  \gamma,\eta\right)
$-fine-shattered by $\mathcal{C}$.
\end{definition}

Clearly $\operatorname*{fine}_{\mathcal{C}}\left(
\gamma,\eta\right) \leq\operatorname*{fat}_{\mathcal{C}}\left(
\gamma\right)  $\ for all $\mathcal{C}$, $\eta$,\ and $\gamma$. \
The following theorem lower-bounds sample complexity in terms of
fine-shattering dimension. \ The proof builds on standard
lower-bound arguments in computational learning theory, such as that
of Ehrenfeucht et al.\ \cite{ehkv}.

\begin{theorem}
\label{finelower}Let $\mathcal{S}$\ be a sample space, let $\mathcal{C}$\ be a
p-concept class over $\mathcal{S}$, and let $\varepsilon,\delta,\gamma,\eta
>0$. \ Then provided $\operatorname*{fine}\nolimits_{\mathcal{C}}\left(
\gamma,\eta\right)  \geq2$ and $\varepsilon,\delta<\frac{1}{4}$, there exists
a distribution $\mathcal{D}$\ over $\mathcal{S}$\ for which the following holds.

\begin{enumerate}
\item[(i)] Suppose%
\[
m\leq\max\left\{  \frac{\operatorname*{fine}\nolimits_{\mathcal{C}}\left(
\gamma,\eta\right)  -1}{64\varepsilon},\frac{1}{4\varepsilon}\ln\frac
{1}{2\delta}\right\}  ,
\]
and let $F\in\mathcal{C}$. \ Then there is no learning algorithm that, given
samples $X=\left(  x_{1},\ldots,x_{m}\right)  $\ drawn independently from
$\mathcal{D}$, as well as real numbers $p_{1},\ldots,p_{m}$\ such that
$\left\vert p_{i}-F\left(  x_{i}\right)  \right\vert \leq\eta$ for all $i$,
outputs a hypothesis $H$ such that%
\[
\Pr_{x\in\mathcal{D}}\left[  \left\vert H\left(  x\right)  -F\left(  x\right)
\right\vert \geq\gamma\right]  \leq\varepsilon
\]
\ with probability at least $1-\delta$\ over $X$.

\item[(ii)] Suppose%
\[
m\leq\max\left\{  \frac{\operatorname*{fine}\nolimits_{\mathcal{C}}\left(
\gamma,\eta\right)  -1}{A\left(  \gamma+\eta\right)  ^{2}\varepsilon},\frac
{1}{4\varepsilon}\ln\frac{1}{2\delta}\right\}
\]
where $A$ is some universal constant, and let $F\in\mathcal{C}$. \ Then there
is no learning algorithm that, given samples $X=\left(  x_{1},\ldots
,x_{m}\right)  $\ drawn independently from $\mathcal{D}$, as well as bits
$b_{1},\ldots,b_{m}$\ such that $\Pr\left[  b_{i}=1\right]  =F\left(
x_{i}\right)  $, outputs a hypothesis $H$ such that%
\[
\Pr_{x\in\mathcal{D}}\left[  \left\vert H\left(  x\right)  -F\left(  x\right)
\right\vert \geq\gamma\right]  \leq\varepsilon
\]
with probability at least $1-\delta$\ over $X$.
\end{enumerate}
\end{theorem}

To finish the proof of Theorem \ref{lower}, the remaining step is to
lower-bound the fine-shattering dimension of $n$-qubit quantum states. \ We
will do so using the following result of Ambainis et al.\ \cite{antv}, which
is basically the converse of Theorem \ref{nayakthm}.

\begin{theorem}
[Ambainis et al.\ \cite{antv}]\label{antvthm}Let $\frac{1}{2}<p<1$, and let
$n$\ and $k$ be positive integers satisfying $n\geq\left(  1-H\left(
p\right)  \right)  k+7\log_{2}k$. \ Then there exist $n$-qubit mixed states
$\left\{  \rho_{y}\right\}  _{y\in\left\{  0,1\right\}  ^{k}}$ and
measurements $E_{1},\ldots,E_{k}$ such that for all $y\in\left\{  0,1\right\}
^{k}$ and $i\in\left\{  1,\ldots,k\right\}  $:

\begin{enumerate}
\item[(i)] if $y_{i}=0$\ then $\operatorname*{Tr}\left(  E_{i}\rho_{y}\right)
\leq1-p$, and

\item[(ii)] if $y_{i}=1$\ then $\operatorname*{Tr}\left(  E_{i}\rho
_{y}\right)  \geq p$.
\end{enumerate}
\end{theorem}

Incidentally, the encoding scheme of Theorem \ref{antvthm}\ is completely
classical, in the sense that the $\rho_{y}$'s and $E_{i}$'s are both diagonal
in the computational basis. \ However, we find it more convenient to state the
result in quantum language.

We will actually need a slight extension of Theorem \ref{antvthm}, which
bounds the $\operatorname*{Tr}\left(  E_{i}\rho_{y}\right)  $'s on both sides
rather than only one.

\begin{theorem}
\label{antvthm2}Let $\frac{1}{2}<p<1$, let $\eta>0$, and let $n$\ and $k$ be
positive integers satisfying $k\leq\frac{2}{\eta}$\ and $n\geq\left(
1-H\left(  p\right)  \right)  k+7\log_{2}\frac{1}{\eta}$. \ Then there exist
$n$-qubit mixed states $\left\{  \rho_{y}\right\}  _{y\in\left\{  0,1\right\}
^{k}}$, and measurements $E_{1},\ldots,E_{k}$, such that for all $y\in\left\{
0,1\right\}  ^{k}$ and $i\in\left\{  1,\ldots,k\right\}  $:

\begin{enumerate}
\item[(i)] if $y_{i}=0$\ then $1-p-\eta\leq\operatorname*{Tr}\left(  E_{i}%
\rho_{y}\right)  \leq1-p$, and

\item[(ii)] if $y_{i}=1$\ then $p\leq\operatorname*{Tr}\left(  E_{i}\rho
_{y}\right)  \leq p+\eta$.
\end{enumerate}
\end{theorem}

As in Section \ref{SC}, let $\mathcal{S}$\ be the set of two-outcome
measurements on $n$ qubits, and let $\mathcal{C}_{n}$\ be class of all
functions $F:\mathcal{S}\rightarrow\left[  0,1\right]  $\ such that $F\left(
E\right)  =\operatorname*{Tr}\left(  E\rho\right)  $ for some $n$-qubit mixed
state $\rho$. \ Then Theorem \ref{antvthm2}\ has the following corollary.

\begin{corollary}
\label{antvcor}For all positive integers $n$ and all $\gamma\geq
\sqrt{n2^{-\left(  n-5\right)  /35}/8}$,%
\[
\operatorname*{fine}\nolimits_{\mathcal{C}_{n}}\left(  \gamma,\frac
{8\gamma^{2}}{n}\right)  \geq\left\lfloor \frac{n}{5\gamma^{2}}\right\rfloor
.
\]

\end{corollary}

Theorem \ref{lower}\ now follows immediately by combining Theorem
\ref{finelower}\ with Corollary \ref{antvcor}.

\section{Appendix: Prediction Problems\label{PRED}}

In this appendix we give a variant of Theorem \ref{qoccam3} that is useful for
prediction (as opposed to learning) problems---and that,\ as a bonus, is
nearly tight. \ As usual, we first give a general upper bound in terms of the
fat-shattering dimension.

\begin{theorem}
\label{alphathm}Let $\mathcal{S}$\ be a sample space, let $\mathcal{C}$\ be a
p-concept class over $\mathcal{S}$,\ and let $\mathcal{D}$\ be a probability
measure over $\mathcal{S}$. \ Fix a p-concept $F\in\mathcal{C}$, as well an
error parameter $\alpha>0$.\ \ Suppose we are given $m$ samples\ $X=\left(
x_{1},\ldots,x_{m}\right)  $\ drawn independently from $\mathcal{D}$, as well
as bits $B=\left(  b_{1},\ldots,b_{m}\right)  $\ such that each $b_{i}$\ is
$1$ with independent probability $F\left(  x_{i}\right)  $. \ Suppose also
that we choose a hypothesis $H\in\mathcal{C}$\ to\ minimize $\sum_{i=1}%
^{m}\left\vert H\left(  x_{i}\right)  -b_{i}\right\vert $. \ Let%
\[
\Delta_{H,F}\left(  x\right)  :=H\left(  x\right)  \left(  1-F\left(
x\right)  \right)  +\left(  1-H\left(  x\right)  \right)  F\left(  x\right)
.
\]
Then there exists a positive constant $K$ such that%
\[
\operatorname*{EX}\limits_{x\in\mathcal{D}}\left[  \Delta_{H,F}\left(
x\right)  \right]  \leq\alpha+\inf_{C\in\mathcal{C}}\operatorname*{EX}%
_{x\in\mathcal{D}}\left[  \Delta_{C,F}\left(  x\right)  \right]
\]
with probability at least $1-\delta$\ over $X$ and $B$, provided that%
\[
m\geq\frac{K}{\alpha^{2}}\left(  \operatorname*{fat}\nolimits_{\mathcal{C}%
}\left(  \frac{\alpha}{10}\right)  \log^{2}\frac{1}{\alpha}+\log\frac
{1}{\delta}\right)  .
\]

\end{theorem}

Theorem \ref{alphathm}\ has the following immediate corollary.

\begin{corollary}
\label{alphacor}Let $\rho$ be an $n$-qubit state, let $\mathcal{D}$ be a
distribution over two-outcome measurements, and let $\mathcal{E}=\left(
E_{1},\ldots,E_{m}\right)  $\ consist of $m$ measurements drawn independently
from $\mathcal{D}$. \ Suppose we are given bits $B=\left(  b_{1},\ldots
,b_{m}\right)  $, where each $b_{i}$\ is $1$ with independent probability
$\operatorname*{Tr}\left(  E_{i}\rho\right)  $. \ Suppose also that we choose
a hypothesis state $\sigma$\ to minimize $\sum_{i=1}^{m}\left\vert
\operatorname*{Tr}\left(  E_{i}\sigma\right)  -b_{i}\right\vert $. \ Let%
\[
\Delta_{\sigma,\rho}\left(  E\right)  :=\operatorname*{Tr}\left(
E\sigma\right)  \left(  1-\operatorname*{Tr}\left(  E\rho\right)  \right)
+\left(  1-\operatorname*{Tr}\left(  E\sigma\right)  \right)
\operatorname*{Tr}\left(  E\rho\right)  .
\]
Then there exists a positive constant $K$ such that%
\[
\operatorname*{EX}\limits_{E\in\mathcal{D}}\left[  \Delta_{\sigma,\rho}\left(
E\right)  \right]  \leq\alpha+\inf_{\varsigma}\operatorname*{EX}%
\limits_{E\in\mathcal{D}}\left[  \Delta_{\varsigma,\rho}\left(  E\right)
\right]
\]
with probability at least $1-\delta$\ over $\mathcal{E}$\ and $B$, provided
that%
\[
m\geq\frac{K}{\alpha^{2}}\left(  \frac{n}{\alpha^{2}}\log^{2}\frac{1}{\alpha
}+\log\frac{1}{\delta}\right)  .
\]

\end{corollary}

Let us describe a simple application of Corollary \ref{alphacor}. \ Given a
two-outcome measurement $E$ and an $n$-qubit state $\rho$,\ let $E\left(
\rho\right)  \in\left\{  0,1\right\}  $\ be the result of applying $E$\ to
$\rho$---that is, $E\left(  \rho\right)  =1$ with probability
$\operatorname*{Tr}\left(  E\rho\right)  $\ and $E\left(  \rho\right)  =0$
otherwise.\ \ Suppose our goal is to output a hypothesis state $\sigma$\ that
maximizes $\Pr_{E\in\mathcal{D}}\left[  E\left(  \sigma\right)  =E\left(
\rho\right)  \right]  $, the \textquotedblleft average probability of
agreement\textquotedblright\ between $\sigma$\ and $\rho$. \ Corollary
\ref{alphacor}\ shows that, by using $O\left(  \frac{n}{\alpha^{4}}\log
^{2}\frac{1}{\alpha}\right)  $ measurements, we can get within an additive
constant $\alpha$\ of the maximum with high probability.

Similarly,\ suppose we are given a measurement $E$ drawn from $\mathcal{D}$,
and want to guess whether $E\left(  \rho\right)  $\ will be $0$ or $1$. \ Here
the maximum success probability is $\frac{1}{2}\operatorname*{EX}%
\nolimits_{E\in\mathcal{D}}\left[  1+\left\vert 2\operatorname*{Tr}\left(
E\rho\right)  -1\right\vert \right]  $, and is obtained by simply guessing $1$
if $\operatorname*{Tr}\left(  E\rho\right)  \geq\frac{1}{2}$, or $0$ if
$\operatorname*{Tr}\left(  E\rho\right)  <\frac{1}{2}$. \ Again, it follows
from Corollary \ref{alphacor} that by using $O\left(  \frac{n}{\alpha^{4}}%
\log^{2}\frac{1}{\alpha}\right)  $\ measurements, we can get within an
additive constant $\alpha$\ of the maximum with high probability.

Using the same arguments as in Appendix \ref{LB}, one can show that Corollary
\ref{alphacor}\ is \textit{tight} up to the $\log^{2}\frac{1}{\alpha}$
term---in particular, that\
\[
m=\Omega\left(  \frac{1}{\alpha^{2}}\left(  \frac{n}{\alpha^{2}}+\log\frac
{1}{\delta}\right)  \right)
\]
measurements are needed. \ We omit the details.

What distinguishes this sort of prediction problem from the learning problems
we have seen before\ is that, as the number of sample measurements $m$ goes to
infinity, we will not necessarily converge to the \textquotedblleft
true\textquotedblright\ state $\rho$. \ One way to see this is that, while
$\rho$\ could be a mixed state, by convexity there is always a pure hypothesis
state $\sigma=\left\vert \psi\right\rangle \left\langle \psi\right\vert
$\ that does as well at the prediction task as any other hypothesis. \ On the
positive side, this means that to \textit{find} such a hypothesis\ given the
measurement results, it suffices to compute the principal eigenvector of a
$2^{n}\times2^{n}$ matrix. \ Unlike for the learning problems, here there is
no need for semidefinite or convex programming.

\section{Appendix: Proofs\label{PROOFS}}

\begin{proof}
[Proof of Corollary \ref{occamcor}]Let $\mathcal{S}$\ be a sample space, let
$\mathcal{C}$\ be a p-concept class over $\mathcal{S}$, and let $\mathcal{D}$
be a probability measure over $\mathcal{S}$. \ Then let $\mathcal{C}^{\ast}%
$\ be the class of p-concepts $G:\mathcal{S}\rightarrow\left[  0,1\right]
$\ for which there exists an $F\in\mathcal{C}$\ such that $\left\vert G\left(
x\right)  -F\left(  x\right)  \right\vert \leq\eta$\ for all $x\in\mathcal{S}%
$. \ Also, fix a p-concept $F\in\mathcal{C}$. \ Suppose we draw $m$ samples
$X=\left(  x_{1},\ldots,x_{m}\right)  $\ independently according to
$\mathcal{D}$, and then choose any hypothesis $H\in\mathcal{C}$\ such that
$\left\vert H\left(  x\right)  -F\left(  x\right)  \right\vert \leq\eta$\ for
all $x\in X$. \ Then there exists a $G\in\mathcal{C}^{\ast}$\ such that
$G\left(  x\right)  =H\left(  x\right)  $ for all $x\in X$. \ This $G$\ is
simply obtained by setting $G\left(  x\right)  :=H\left(  x\right)  $\ if
$x\in X$\ and $G\left(  x\right)  :=F\left(  x\right)  $\ otherwise.

So by Theorem \ref{occam2}, provided that%
\[
m\geq\frac{K}{\alpha^{2}}\left(  \operatorname*{fat}\nolimits_{\mathcal{C}%
^{\ast}}\left(  \frac{\alpha}{5}\right)  \log^{2}\frac{1}{\alpha}+\log\frac
{1}{\delta}\right)  ,
\]
we have%
\[
\operatorname*{EX}_{x\in\mathcal{D}}\left[  \left\vert H\left(  x\right)
-G\left(  x\right)  \right\vert \right]  \leq\alpha+\inf_{C\in\mathcal{C}%
^{\ast}}\operatorname*{EX}_{x\in\mathcal{D}}\left[  \left\vert C\left(
x\right)  -G\left(  x\right)  \right\vert \right]  =\alpha
\]
with probability at least $1-\delta$ over $X$. \ Here we have used the fact
that $G\in\mathcal{C}^{\ast}$\ and hence%
\[
\inf_{C\in\mathcal{C}^{\ast}}\operatorname*{EX}_{x\in\mathcal{D}}\left[
\left\vert C\left(  x\right)  -G\left(  x\right)  \right\vert \right]  =0.
\]
Setting $\alpha:=\frac{6\gamma}{7}\varepsilon$, this implies by Markov's
inequality that%
\[
\Pr_{x\in\mathcal{D}}\left[  \left\vert H\left(  x\right)  -G\left(  x\right)
\right\vert >\frac{6\gamma}{7}\right]  \leq\varepsilon,
\]
and therefore%
\[
\Pr_{x\in\mathcal{D}}\left[  \left\vert H\left(  x\right)  -F\left(  x\right)
\right\vert >\frac{6\gamma}{7}+\eta\right]  \leq\varepsilon.
\]
Since $\eta\leq\frac{\gamma\varepsilon}{7}\leq\frac{\gamma}{7}$, the above
implies that%
\[
\Pr_{x\in\mathcal{D}}\left[  \left\vert H\left(  x\right)  -F\left(  x\right)
\right\vert >\gamma\right]  \leq\varepsilon
\]
as desired.

Next we claim that $\operatorname*{fat}\nolimits_{\mathcal{C}^{\ast}}\left(
\alpha\right)  \leq\operatorname*{fat}\nolimits_{\mathcal{C}}\left(
\alpha-\eta\right)  $. \ The reason is simply that, if a given set $\alpha
$-fat-shatters $\mathcal{C}^{\ast}$,\ then it must also $\left(  \alpha
-\eta\right)  $-fat-shatter $\mathcal{C}$ by the triangle inequality.

Putting it all together, we have%
\[
\operatorname*{fat}\nolimits_{\mathcal{C}^{\ast}}\left(  \frac{\alpha}%
{5}\right)  \leq\operatorname*{fat}\nolimits_{\mathcal{C}}\left(  \frac
{\alpha}{5}-\eta\right)  \leq\operatorname*{fat}\nolimits_{\mathcal{C}}\left(
\frac{6\gamma\varepsilon/7}{5}-\frac{\gamma\varepsilon}{7}\right)
=\operatorname*{fat}\nolimits_{\mathcal{C}}\left(  \frac{\gamma\varepsilon
}{35}\right)  ,
\]
and hence%
\[
m\geq\frac{K}{\alpha^{2}}\left(  \operatorname*{fat}\nolimits_{\mathcal{C}%
}\left(  \frac{\gamma\varepsilon}{35}\right)  \log^{2}\frac{1}{\alpha}%
+\log\frac{1}{\delta}\right)  =\frac{K}{\left(  6\gamma\varepsilon/7\right)
^{2}}\left(  \operatorname*{fat}\nolimits_{\mathcal{C}}\left(  \frac
{\gamma\varepsilon}{35}\right)  \log^{2}\frac{1}{6\gamma\varepsilon/7}%
+\log\frac{1}{\delta}\right)
\]
samples suffice.
\end{proof}

\begin{proof}
[Proof of Theorem \ref{nayakthm2}]Suppose there exists such an encoding scheme
with $n/\gamma^{2}=o\left(  k\right)  $. \ Then consider an amplified scheme,
where each string $y\in\left\{  0,1\right\}  ^{k}$ is encoded by the tensor
product state $\rho_{y}^{\otimes\ell}$. \ Here we set $\ell:=\left\lceil
c/\gamma^{2}\right\rceil $\ for some $c>0$. \ Also, for all $i\in\left\{
1,\ldots,k\right\}  $, let $E_{i}^{\ast}$\ be an amplified measurement that
applies $E_{i}$\ to each of the $\ell$\ copies of $\rho_{y}$, and accepts if
and only if at least $\alpha_{i}\ell$\ of the $E_{i}$'s do. \ Then provided we
choose $c$ sufficiently large, it is easy to show by a Chernoff bound that for
all $y$\ and $i$,

\begin{enumerate}
\item[(i)] if $y_{i}=0$\ then $\operatorname*{Tr}\left(  E_{i}^{\ast}\rho
_{y}^{\otimes\ell}\right)  \leq\frac{1}{3}$, and

\item[(ii)] if $y_{i}=1$\ then $\operatorname*{Tr}\left(  E_{i}^{\ast}\rho
_{y}^{\otimes\ell}\right)  \geq\frac{2}{3}$.
\end{enumerate}

So to avoid contradicting Theorem \ref{nayakthm}, we need
$n\ell\geq\left( 1-H\left(  \frac{1}{3}\right)  \right)  k$. \ But
this implies that $n/\gamma^{2}=\Omega\left(  k\right)
$.\footnote{If we care about optimizing the constant under the
$\Omega\left(  k\right)  $, then we are better off avoiding
amplication and instead proving Theorem \ref{nayakthm2}\ directly
using the techniques of Ambainis et al.\ \cite{antv}. \ Doing so, we
obtain $n/\gamma^{2}\geq2k/\ln2$.}
\end{proof}

\begin{proof}
[Proof of Theorem \ref{ccthm}]Let $f:\mathcal{Z}\rightarrow\left\{
0,1\right\}  $ be a Boolean function with $\mathcal{Z}\subseteq\left\{
0,1\right\}  ^{N}\times\left\{  0,1\right\}  ^{M}$. \ Fix Alice's input
$x\in\left\{  0,1\right\}  ^{N}$, and let $\mathcal{Z}_{x}$\ be the set of all
$y\in\left\{  0,1\right\}  ^{M}$\ such that $\left(  x,y\right)
\in\mathcal{Z}$. \ By Yao's minimax principle, to give a randomized protocol
that errs with probability at most $\frac{1}{3}$\ for all $y\in\mathcal{Z}%
_{x}$, it is enough, for any fixed\textit{ }probability distribution
$\mathcal{D}$\ over $\mathcal{Z}_{x}$, to give a randomized protocol that errs
with probability at most $\frac{1}{3}$\ over $y$ drawn from $\mathcal{D}%
$.\footnote{Indeed, it suffices to give a \textit{deterministic} protocol that
errs with probability at most $\frac{1}{3}$\ over $y$ drawn from $\mathcal{D}%
$, a fact we will not need.}

So let $\mathcal{D}$ be such a distribution; then the randomized protocol is
as follows. \ First Alice chooses $k$ inputs $y_{1},\ldots,y_{k}%
$\ independently from $\mathcal{D}$, where $k=O\left(  \operatorname*{Q}%
\nolimits^{1}\left(  f\right)  \right)  $. \ She then sends Bob $y_{1}%
,\ldots,y_{k}$, together with $f\left(  x,y_{i}\right)  $\ for all
$i\in\left\{  1,\ldots,k\right\}  $. \ Clearly this message requires only
$O\left(  M\operatorname*{Q}\nolimits^{1}\left(  f\right)  \right)  $
classical bits. \ We need to show that it lets Bob evaluate $f\left(
x,y\right)  $, with high probability over $y$ drawn from $\mathcal{D}$.

By amplification, we can assume Bob errs with probability at most $\eta$ for
any fixed constant $\eta>0$. \ We will take $\eta=\frac{1}{100}$. \ Also, in
the quantum protocol for $f$, let $\rho_{x}$\ be the $\operatorname*{Q}%
\nolimits^{1}\left(  f\right)  $-qubit mixed state that Alice would send given
input $x$, and let $E_{y}$\ be the measurement that Bob would apply given
input $y$. \ Then $\operatorname*{Tr}\left(  E_{y}\rho_{x}\right)  \geq1-\eta
$\ if $f\left(  x,y\right)  =1$,\ while $\operatorname*{Tr}\left(  E_{y}%
\rho_{x}\right)  \leq\eta$\ if $f\left(  x,y\right)  =0$.

Given Alice's classical message, first Bob finds a $\operatorname*{Q}%
\nolimits^{1}\left(  f\right)  $-qubit state $\sigma$\ such that $\left\vert
\operatorname*{Tr}\left(  E_{y_{i}}\sigma\right)  -f\left(  x,y_{i}\right)
\right\vert \leq\eta$\ for all $i\in\left\{  1,\ldots,k\right\}  $.
\ Certainly such a state\ exists (for take $\sigma=\rho_{x}$), and Bob can
find it by searching exhaustively for its classical description. \ If there
are multiple such states, then Bob chooses one in some arbitrary deterministic
way (for example, by lexicographic ordering). \ Note that we then have
$\left\vert \operatorname*{Tr}\left(  E_{y_{i}}\sigma\right)
-\operatorname*{Tr}\left(  E_{y_{i}}\rho_{x}\right)  \right\vert \leq\eta
$\ for all $i\in\left\{  1,\ldots,k\right\}  $ as well. \ Finally Bob outputs
$f\left(  x,y\right)  =1$\ if $\operatorname*{Tr}\left(  E_{y}\sigma\right)
\geq\frac{1}{2}$,\ or $f\left(  x,y\right)  =0$\ if $\operatorname*{Tr}\left(
E_{y}\sigma\right)  <\frac{1}{2}$.

Set $\varepsilon=\delta=\frac{1}{6}$ and $\gamma=0.42$,\ so that
$\gamma\varepsilon=7\eta$. \ Then by Theorem \ref{qoccam},%
\[
\Pr_{y\in\mathcal{D}}\left[  \left\vert \operatorname*{Tr}\left(  E_{y}%
\sigma\right)  -\operatorname*{Tr}\left(  E_{y}\rho_{x}\right)  \right\vert
>\gamma\right]  >\varepsilon
\]
with probability at most $\delta$\ over Alice's classical message, provided
that%
\[
k=\Omega\left(  \frac{1}{\gamma^{2}\varepsilon^{2}}\left(  \frac
{\operatorname*{Q}\nolimits^{1}\left(  f\right)  }{\gamma^{2}\varepsilon^{2}%
}\log^{2}\frac{1}{\gamma\varepsilon}+\log\frac{1}{\delta}\right)  \right)  .
\]
So in particular, there exist constants $A,B$\ such that if $k\geq
A\operatorname*{Q}\nolimits^{1}\left(  f\right)  +B$, then%
\[
\Pr_{y\in\mathcal{D}}\left[  \left\vert \operatorname*{Tr}\left(  E_{y}%
\sigma\right)  -f\left(  x,y\right)  \right\vert >\gamma+\eta\right]
>\varepsilon
\]
with probability most $\delta$. \ Since $\gamma+\eta<\frac{1}{2}$, it follows
that Bob's classical strategy will fail with probability at most
$\varepsilon+\delta=\frac{1}{3}$ over $y$ drawn from $\mathcal{D}$.
\end{proof}

\begin{proof}
[Proof of Theorem \ref{ypthm}]\quad

\begin{enumerate}
\item[(i)] Similar to the proof that $\mathsf{BPP}\subset\mathsf{P/poly}$.
\ Given a $\mathsf{ZPP}$\ machine $M$, first amplify $M$ so that its failure
probability on any input of length $n$\ is at most $2^{-2n}$. \ Then by a
counting argument, there exists a single random string $r_{n}$ that causes $M$
to succeed on all $2^{n}$\ inputs simultaneously. \ Use that $r_{n}$\ as the
$\mathsf{YP}$\ machine's advice.

\item[(ii)] $\mathsf{YE}\subseteq\mathsf{NE}\cap\mathsf{coNE}$\ is immediate.
\ For $\mathsf{NE}\cap\mathsf{coNE}\subseteq\mathsf{YE}$, first concatenate
the $\mathsf{NE}$\ and $\mathsf{coNE}$\ witnesses for all $2^{n}$\ inputs of
length $n$, then use the resulting string (of length $2^{O\left(  n\right)  }%
$) as the $\mathsf{YE}$\ machine's advice.

\item[(iii)] If $\mathsf{P}=\mathsf{YP}$\ then $\mathsf{E}=\mathsf{YE}$\ by
padding. \ Hence $\mathsf{E}=\mathsf{NE}\cap\mathsf{coNE}$ by part (ii).

\item[(iv)] Let $M$ be a $\mathsf{YP}$\ machine, and let $y_{n}$\ be the
lexicographically first advice string that causes $M$\ to succeed on all
$2^{n}$\ inputs of length $n$. \ Consider the following computational problem:
\textit{given integers }$\left\langle n,i\right\rangle $\textit{ encoded in
binary,\ compute the }$i^{th}$\textit{\ bit of }$y_{n}$. \ We claim that this
problem is in $\mathsf{NE}^{\mathsf{NP}^{\mathsf{NP}}}$. \ For an
$\mathsf{NE}^{\mathsf{NP}^{\mathsf{NP}}}$\ machine can first guess $y_{n}$,
then check that it works for all $x\in\left\{  0,1\right\}  ^{n}$\ using
$\mathsf{NP}$\ queries, then check that no lexicographically earlier string
\textit{also} works using $\mathsf{NP}^{\mathsf{NP}}$\ queries, and finally
return the $i^{th}$\ bit of $y_{n}$. \ So if $\mathsf{E}=\mathsf{NE}%
^{\mathsf{NP}^{\mathsf{NP}}}$, then the problem is in $\mathsf{E}$, which
means that an $\mathsf{E}$ machine can recover $y_{n}$ itself by simply
looping over all $i$. \ So if $n$ and $i$ take only logarithmically many bits
to specify, then a $\mathsf{P}$\ machine can recover $y_{n}$. \ Hence
$\mathsf{P}=\mathsf{YP}$.
\end{enumerate}
\end{proof}

\begin{proof}
[Proof of Lemma \ref{thelem}]The procedure $Q$ is given by the following
pseudocode:\bigskip

\texttt{\qquad Let }$\rho:=\rho_{0}$

\texttt{\qquad Choose }$t\in\left\{  1,\ldots,T\right\}  $\texttt{\ uniformly
at random}

\texttt{\qquad For }$u:=1$\texttt{\ to }$t$

\texttt{\qquad\qquad Choose }$i\in\left\{  1,\ldots,m\right\}  $\texttt{
uniformly at random}

\texttt{\qquad\qquad Apply }$E_{i}$\texttt{\ to }$\rho$

\texttt{\qquad\qquad If }$E_{i}$\texttt{ rejects, return "FAILURE"\ and halt}

\texttt{\qquad Next }$u$

\texttt{\qquad Return "SUCCESS"\ and output }$\sigma:=\rho\bigskip$

Property (ii) follows immediately from Proposition \ref{qunion}. \ For
property (iii), let $\rho_{u}$\ be the state of $\rho$\ immediately after the
$u^{th}$\ iteration, conditioned on iterations $1,\ldots,u$\ all succeeding.
\ Also, let $\beta_{u}:=\max_{i}\left\{  1-\operatorname*{Tr}\left(  E_{i}%
\rho_{u}\right)  \right\}  $. \ Then $Q$ fails in the $\left(  u+1\right)
^{st}$\ iteration with probability at least $\beta_{u}/m$, conditioned on
succeeding in iterations $1,\ldots,u$. \ So letting $p_{t}$\ be the
probability that $Q$ completes all $t$ iterations, we have%
\[
p_{t}\leq\left(  1-\frac{\beta_{0}}{m}\right)  \cdots\left(  1-\frac
{\beta_{t-1}}{m}\right)  .
\]
Hence, letting $z>0$ be a parameter to be determined later,%
\begin{align*}
\sum_{t~:~\beta_{t}>z}p_{t} &  \leq\sum_{t~:~\beta_{t}>z}\left(  1-\frac
{\beta_{0}}{m}\right)  \cdots\left(  1-\frac{\beta_{t-1}}{m}\right)  \\
&  \leq\sum_{t~:~\beta_{t}>z}~\prod_{u<t~:~\beta_{u}>z}\left(  1-\frac
{\beta_{u}}{m}\right)  \\
&  \leq\sum_{t=0}^{\infty}\left(  1-\frac{z}{m}\right)  ^{t}\\
&  =\frac{m}{z}.
\end{align*}
Also, by the assumption that $Q$ succeeds\ with probability at least $\lambda
$, we have $\frac{1}{T}\sum_{t}p_{t}\geq\lambda$. \ So for all $i$,%
\begin{align*}
1-\operatorname*{Tr}\left(  E_{i}\sigma\right)   &  =\frac{\sum_{t}%
p_{t}\left(  1-\operatorname*{Tr}\left(  E_{i}\rho_{t}\right)  \right)  }%
{\sum_{t}p_{t}}\\
&  =\frac{\sum_{t~:~\beta_{t}\leq z}p_{t}\left(  1-\operatorname*{Tr}\left(
E_{i}\rho_{t}\right)  \right)  }{\sum_{t}p_{t}}+\frac{\sum_{t~:~\beta_{t}%
>z}p_{t}\left(  1-\operatorname*{Tr}\left(  E_{i}\rho_{t}\right)  \right)
}{\sum_{t}p_{t}}\\
&  \leq\frac{\sum_{t~:~\beta_{t}\leq z}p_{t}\beta_{t}}{\sum_{t}p_{t}}%
+\frac{m/z}{\sum_{t}p_{t}}\\
&  \leq z+\frac{m/z}{\lambda T}.
\end{align*}
The last step is to set $z:=\sqrt{\frac{m}{\lambda T}}$, thereby obtaining the
optimal lower bound%
\[
\operatorname*{Tr}\left(  E_{i}\sigma\right)  \geq1-2\sqrt{\frac{m}{\lambda
T}}.
\]

\end{proof}

\begin{proof}
[Proof of Theorem \ref{advthm}]Fix a distributional problem $\left(
L,\left\{  \mathcal{D}_{n}\right\}  \right)  \in\mathsf{HeurBQP/qpoly}$.
\ Then there exists a polynomial-time quantum algorithm $A$ such that for all
$n$ and $\varepsilon>0$,\ there exists a state $\left\vert \psi_{n,\varepsilon
}\right\rangle $\ of size $q=O\left(  \operatorname*{poly}\left(
n,1/\varepsilon\right)  \right)  $ such that%
\[
\Pr_{x\in\mathcal{D}_{n}}\left[  P_{A}^{L\left(  x\right)  }\left(  \left\vert
x\right\rangle \left\vert \psi_{n,\varepsilon}\right\rangle \right)  \geq
\frac{2}{3}\right]  \geq1-\varepsilon.
\]
Let $\mathcal{D}_{n}^{\ast}$\ be the distribution obtained by starting from
$\mathcal{D}_{n}$ and then conditioning on $P_{A}^{L\left(  x\right)  }\left(
\left\vert x\right\rangle \left\vert \psi_{n,\varepsilon}\right\rangle
\right)  \geq\frac{2}{3}$. \ Then our goal will be to construct a
polynomial-time verification procedure $V$ such that, for all $n$ and
$\varepsilon>0$,\ there exists an advice string $a_{n,\varepsilon}\in\left\{
0,1\right\}  ^{\operatorname*{poly}\left(  n,1/\varepsilon\right)  }$\ for
which the following holds.

\begin{itemize}
\item There exists a state $\left\vert \varphi_{n,\varepsilon}\right\rangle
\in\mathcal{H}_{2}^{\otimes\operatorname*{poly}\left(  n,1/\varepsilon\right)
}$\ such that%
\[
\Pr_{x\in\mathcal{D}_{n}^{\ast}}\left[  P_{V}^{L\left(  x\right)  }\left(
\left\vert x\right\rangle \left\vert \varphi_{n,\varepsilon}\right\rangle
\left\vert a_{n,\varepsilon}\right\rangle \right)  \geq\frac{2}{3}\right]
\geq1-\varepsilon.
\]

\item The probability over $x\in\mathcal{D}_{n}^{\ast}$\ that there exists a
state $\left\vert \varphi\right\rangle $\ such that $P_{V}^{1-L\left(
x\right)  }\left(  \left\vert x\right\rangle \left\vert \varphi\right\rangle
\left\vert a_{n,\varepsilon}\right\rangle \right)  \geq\frac{1}{3}$\ is at
most $\varepsilon$.
\end{itemize}

If $V$ succeeds with probability at least $1-\varepsilon$\ over $x\in
\mathcal{D}_{n}^{\ast}$, then by the union bound it succeeds with probability
at least $1-2\varepsilon$\ over $x\in\mathcal{D}_{n}$. \ Clearly this suffices
to prove the theorem.

As a preliminary step, let us replace $A$\ by an amplified algorithm $A^{\ast
}$, which takes $\left\vert \psi_{n,\varepsilon}\right\rangle ^{\otimes\ell}%
$\ as advice and returns the majority answer among $\ell$\ invocations of $A$.
\ Here $\ell$\ is a parameter to be determined later. \ By a Chernoff bound,%
\[
\Pr_{x\in\mathcal{D}_{n}}\left[  P_{A^{\ast}}^{L\left(  x\right)  }\left(
\left\vert x\right\rangle \left\vert \psi_{n,\varepsilon}\right\rangle
^{\otimes\ell}\right)  \geq1-e^{-\ell/18}\right]  \geq1-\varepsilon.
\]

We now describe the verifier $V$. \ The verifier receives three objects as input:

\begin{itemize}
\item \textbf{An input }$x\in\left\{  0,1\right\}  ^{n}$\textbf{.}

\item \textbf{An untrusted quantum advice state }$\left\vert \varphi
_{0}\right\rangle $\textbf{.} \ This $\left\vert \varphi_{0}\right\rangle
$\ is divided into $\ell$\ registers, each with $q$ qubits. \ The state that
the verifier \textit{expects} to receive is $\left\vert \varphi_{0}%
\right\rangle =\left\vert \psi_{n,\varepsilon}\right\rangle ^{\otimes\ell}$.

\item \textbf{A trusted classical advice string }$a_{n,\varepsilon}$\textbf{.}
\ This $a_{n,\varepsilon}$\ consists of $m$ test inputs $x_{1},\ldots,x_{m}%
\in\left\{  0,1\right\}  ^{n}$, together with $L\left(  x_{i}\right)  $\ for
$i\in\left\{  1,\ldots,m\right\}  $. \ Here $m$\ is a parameter to be
determined later.
\end{itemize}

Given these objects, $V$ does the following,\ where $T$ is another parameter
to be determined later.\bigskip

\qquad\textbf{Phase 1: Verify }$\left\vert \varphi_{0}\right\rangle $

\qquad\texttt{Let }$\left\vert \varphi\right\rangle :=\left\vert \varphi
_{0}\right\rangle $

\qquad\texttt{Choose }$t\in\left\{  1,\ldots,T\right\}  $\texttt{\ uniformly
at random}

\qquad\texttt{For }$u:=1$\texttt{\ to }$t$

\qquad\qquad\texttt{Choose }$i\in\left\{  1,\ldots,m\right\}  $\texttt{
uniformly at random}

\qquad\qquad\texttt{Simulate }$A^{\ast}\left(  \left\vert x_{i}\right\rangle
\left\vert \varphi\right\rangle \right)  $

\qquad\qquad\texttt{If }$A^{\ast}$\texttt{ outputs }$1-L\left(  x_{i}\right)
$\texttt{, output "don't know"\ and halt}

\qquad\texttt{Next }$u\bigskip$

\qquad\textbf{Phase 2: Decide whether }$x\in L$

\qquad\texttt{Simulate }$A^{\ast}\left(  \left\vert x\right\rangle \left\vert
\varphi\right\rangle \right)  $

\qquad\texttt{Accept if }$A^{\ast}$\texttt{\ outputs }$1$\texttt{; reject
otherwise\bigskip}

It suffices to show that there exists a choice of test inputs $x_{1}%
,\ldots,x_{m}$, as well as parameters $\ell$, $m$, and $T$, for which the
following holds.

\begin{enumerate}
\item[(a)] If $\left\vert \varphi_{0}\right\rangle =\left\vert \psi
_{n,\varepsilon}\right\rangle ^{\otimes\ell}$, then Phase 1 succeeds with
probability at least $\frac{5}{6}$.

\item[(b)] If Phase 1 succeeds with probability at least $\frac{1}{3}$, then
conditioned on its succeeding, $P_{A^{\ast}}^{L\left(  x\right)  }\left(
\left\vert x_{i}\right\rangle \left\vert \varphi\right\rangle \right)
\geq\frac{17}{18}$\ for all $i\in\left\{  1,\ldots,m\right\}  $.

\item[(c)] If $P_{A^{\ast}}^{L\left(  x\right)  }\left(  \left\vert
x_{i}\right\rangle \left\vert \varphi\right\rangle \right)  \geq\frac{17}{18}%
$\ for all $i\in\left\{  1,\ldots,m\right\}  $,\ then%
\[
\Pr_{x\in\mathcal{D}_{n}^{\ast}}\left[  P_{A^{\ast}}^{L\left(  x\right)
}\left(  \left\vert x\right\rangle \left\vert \varphi\right\rangle \right)
\geq\frac{5}{6}\right]  \geq1-\varepsilon.
\]

\end{enumerate}

For conditions (a)-(c) ensure that the following holds with probability at
least $1-\varepsilon$\ over $x\in\mathcal{D}_{n}^{\ast}$. \ First, if
$\left\vert \varphi_{0}\right\rangle =\left\vert \psi_{n,\varepsilon
}\right\rangle ^{\otimes\ell}$, then%
\[
P_{V}^{L\left(  x\right)  }\left(  \left\vert x\right\rangle \left\vert
\varphi_{0}\right\rangle \left\vert a_{n,\varepsilon}\right\rangle \right)
\geq\frac{5}{6}-\frac{1}{6}=\frac{2}{3}%
\]
by the union bound. \ Here $\frac{1}{6}$\ is the maximum probability of
failure in Phase 1, while $\frac{5}{6}$ is the minimum probability of success
in Phase 2. \ Second, for all $\left\vert \varphi_{0}\right\rangle $, either
Phase 1 succeeds with probability less than $\frac{1}{3}$, or else Phase 2
succeeds with probability at least $\frac{5}{6}$. \ Hence%
\[
P_{V}^{1-L\left(  x\right)  }\left(  \left\vert x\right\rangle \left\vert
\varphi_{0}\right\rangle \left\vert a_{n,\varepsilon}\right\rangle \right)
\leq\max\left\{  \frac{1}{3},\frac{1}{6}\right\}  =\frac{1}{3}.
\]
Therefore $V$\ is a valid\ $\mathsf{HeurYQP/poly}$\ verifier as desired.

Set%
\begin{align*}
m &  :=K\frac{q}{\varepsilon}\log^{3}\frac{q}{\varepsilon},\\
\ell &  :=100+9\ln m,\\
T &  :=3888m,
\end{align*}
where $K>0$\ is a sufficiently large constant and $q$ is the number of qubits
of $\left\vert \psi_{n,\varepsilon}\right\rangle $. \ Also, form the advice
string $a_{n,\varepsilon}$\ by choosing $x_{1},\ldots,x_{m}$\ independently
from $\mathcal{D}_{n}^{\ast}$. \ We will show that conditions (a)-(c) all hold
with high probability over the choice of $x_{1},\ldots,x_{m}$---and hence,
that there certainly \textit{exists} a choice of $x_{1},\ldots,x_{m}$\ for
which they hold.

To prove (a), we appeal to part (ii) of Lemma \ref{thelem}. \ Setting
$\epsilon:=e^{-\ell/18}$, we have $P_{A^{\ast}}^{L\left(  x\right)  }\left(
\left\vert x_{i}\right\rangle \left\vert \psi_{n,\varepsilon}\right\rangle
^{\otimes\ell}\right)  \geq1-\epsilon$\ for all $i\in\left\{  1,\ldots
,m\right\}  $. \ Therefore Phase 1 succeeds with probability at least%
\[
1-T\sqrt{\epsilon}=1-3888m\cdot e^{-\ell/9}\geq\frac{5}{6}.
\]

To prove (b), we appeal to part (iii) of Lemma \ref{thelem}. \ Set
$\lambda:=\frac{1}{3}$. \ Then if Phase 1 succeeds with probability at least
$\lambda$, for all $i$\ we have%
\[
P_{A^{\ast}}^{L\left(  x\right)  }\left(  \left\vert x_{i}\right\rangle
\left\vert \varphi\right\rangle \right)  \geq1-2\sqrt{\frac{m}{\lambda T}%
}=1-2\sqrt{\frac{3m}{3888m}}=\frac{17}{18}.
\]

Finally, to prove (c), we appeal to Theorem \ref{qoccam2}. \ Set $\eta
:=\frac{1}{18}$. \ Then for all $i$ we have%
\[
P_{A^{\ast}}^{L\left(  x\right)  }\left(  \left\vert x_{i}\right\rangle
\left\vert \varphi\right\rangle \right)  \geq\frac{17}{18}=1-\eta,
\]
and also%
\[
P_{A^{\ast}}^{L\left(  x\right)  }\left(  \left\vert x_{i}\right\rangle
\left\vert \psi_{n,\varepsilon}\right\rangle ^{\otimes\ell}\right)
\geq1-e^{-\ell/18}>1-\eta.
\]
Hence%
\[
\left\vert P_{A^{\ast}}^{L\left(  x\right)  }\left(  \left\vert x_{i}%
\right\rangle \left\vert \varphi\right\rangle \right)  -P_{A^{\ast}}^{L\left(
x\right)  }\left(  \left\vert x_{i}\right\rangle \left\vert \psi
_{n,\varepsilon}\right\rangle ^{\otimes\ell}\right)  \right\vert \leq\eta.
\]
Now set $\gamma:=\frac{1}{9}$ and $\delta:=\frac{1}{3}$. \ Then $\gamma>\eta
$\ and%
\begin{align*}
m &  =\Omega\left(  \frac{q}{\varepsilon}\log^{3}\frac{q}{\varepsilon}\right)
\\
&  =\Omega\left(  \frac{q\ell}{\varepsilon}\log^{2}\frac{q\ell}{\varepsilon
}\right)  \\
&  =\Omega\left(  \frac{1}{\varepsilon}\left(  \frac{q\ell}{\left(
\gamma-\eta\right)  ^{2}}\log^{2}\frac{q\ell}{\left(  \gamma-\eta\right)
\varepsilon}+\log\frac{1}{\delta}\right)  \right)  .
\end{align*}
So Theorem \ref{qoccam2}\ implies that%
\[
\Pr_{x\in\mathcal{D}_{n}^{\ast}}\left[  \left\vert P_{A^{\ast}}^{L\left(
x\right)  }\left(  \left\vert x\right\rangle \left\vert \varphi\right\rangle
\right)  -P_{A^{\ast}}^{L\left(  x\right)  }\left(  \left\vert x\right\rangle
\left\vert \psi_{n,\varepsilon}\right\rangle ^{\otimes\ell}\right)
\right\vert >\gamma\right]  \leq\varepsilon
\]
and hence%
\[
\Pr_{x\in\mathcal{D}_{n}^{\ast}}\left[  P_{A^{\ast}}^{L\left(  x\right)
}\left(  \left\vert x\right\rangle \left\vert \varphi\right\rangle \right)
<\frac{5}{6}\right]  \leq\varepsilon
\]
with probability at least $1-\delta$\ over the choice of $a_{n,\varepsilon}$.
\ Here we have used the facts that%
\[
P_{A^{\ast}}^{L\left(  x\right)  }\left(  \left\vert x\right\rangle \left\vert
\psi_{n,\varepsilon}\right\rangle ^{\otimes\ell}\right)  \geq1-\eta
\]
and that $\eta+\gamma=\frac{1}{18}+\frac{1}{9}=\frac{1}{6}$.
\end{proof}

\begin{proof}
[Proof of Theorem \ref{alonthm}]Let $\mathcal{D}^{\ast}$\ be the distribution
over $\left(  x,b\right)  \in\mathcal{S}\times\left\{  0,1\right\}
$\ obtained by first drawing $x$ from $\mathcal{D}$, and then setting $b=1$
with probability $F\left(  x\right)  $ and $b=0$ with probability $1-F\left(
x\right)  $. \ Then we can imagine that each $\left(  x_{i},b_{i}\right)  $
was drawn from $\mathcal{D}^{\ast}$. \ Also, given a hypothesis $H\in
\mathcal{C}$, let $H^{\ast}:\mathcal{S}\times\left\{  0,1\right\}
\rightarrow\left[  0,1\right]  $\ be the quadratic loss function\ defined by
$H^{\ast}\left(  x,0\right)  =H\left(  x\right)  ^{2}$\ and $H^{\ast}\left(
x,1\right)  =\left(  1-H\left(  x\right)  \right)  ^{2}$. \ Then let
$\mathcal{C}^{\ast}$\ be the p-concept class consisting of $H^{\ast}$\ for all
$H\in\mathcal{C}$.

Call $H^{\ast}\in\mathcal{C}^{\ast}$\ an \textquotedblleft$\alpha
$-good\textquotedblright\ function if%
\[
\operatorname*{EX}_{\left(  x,b\right)  \in\mathcal{D}^{\ast}}\left[  H^{\ast
}\left(  x,b\right)  \right]  \leq\alpha+\inf_{C^{\ast}\in\mathcal{C}^{\ast}%
}\operatorname*{EX}_{\left(  x,b\right)  \in\mathcal{D}^{\ast}}\left[
C^{\ast}\left(  x,b\right)  \right]  .
\]
Notice that%
\begin{align*}
\operatorname*{EX}_{\left(  x,b\right)  \in\mathcal{D}^{\ast}}\left[  H^{\ast
}\left(  x,b\right)  \right]   &  =\operatorname*{EX}_{x\in\mathcal{D}}\left[
\left(  1-F\left(  x\right)  \right)  H\left(  x\right)  ^{2}+F\left(
x\right)  \left(  1-H\left(  x\right)  \right)  ^{2}\right]  \\
&  =\operatorname*{EX}_{x\in\mathcal{D}}\left[  \left(  H\left(  x\right)
-F\left(  x\right)  \right)  ^{2}+F\left(  x\right)  -F\left(  x\right)
^{2}\right]  .
\end{align*}
Therefore, if $H^{\ast}$\ is $\alpha$-good\ then%
\[
\operatorname*{EX}_{x\in\mathcal{D}}\left[  \left(  H\left(  x\right)
-F\left(  x\right)  \right)  ^{2}\right]  \leq\alpha+\inf_{C\in\mathcal{C}%
}\operatorname*{EX}_{x\in\mathcal{D}}\left[  \left(  C\left(  x\right)
-F\left(  x\right)  \right)  ^{2}\right]  .
\]
Since%
\[
\inf_{C\in\mathcal{C}}\operatorname*{EX}_{x\in\mathcal{D}}\left[  \left(
C\left(  x\right)  -F\left(  x\right)  \right)  ^{2}\right]  =0,
\]
this implies in particular that%
\[
\operatorname*{EX}_{x\in\mathcal{D}}\left[  \left(  H\left(  x\right)
-F\left(  x\right)  \right)  ^{2}\right]  \leq\alpha.
\]
If we set $\alpha:=\gamma^{2}\varepsilon$, then the above implies by Markov's
inequality that%
\[
\Pr_{x\in\mathcal{D}}\left[  \left\vert H\left(  x\right)  -F\left(  x\right)
\right\vert >\gamma\right]  \leq\varepsilon
\]
as desired.

Now suppose we are given $m$ samples $\left(  x_{1},b_{1}\right)
,\ldots,\left(  x_{m},b_{m}\right)  $ drawn independently from $\mathcal{D}%
^{\ast}$. \ Also, let $Z\left(  x,b\right)  =0$\ be the identically-zero
function. \ Then Theorem \ref{occam2} implies that, if we choose $H^{\ast}%
\in\mathcal{C}^{\ast}$ to minimize%
\[
\sum_{i=1}^{m}\left\vert H^{\ast}\left(  x_{i},b_{i}\right)  -Z\left(
x_{i},b_{i}\right)  \right\vert =\sum_{i=1}^{m}H^{\ast}\left(  x_{i}%
,b_{i}\right)  =\sum_{i=1}^{m}\left(  H\left(  x_{i}\right)  -b_{i}\right)
^{2},
\]
then $H^{\ast}$ will be $\alpha$-good\ with probability at least $1-\delta$,
provided that%
\begin{align*}
m &  =O\left(  \frac{1}{\alpha^{2}}\left(  \operatorname*{fat}%
\nolimits_{\mathcal{C}^{\ast}}\left(  \frac{\alpha}{5}\right)  \log^{2}%
\frac{1}{\alpha}+\log\frac{1}{\delta}\right)  \right)  \\
&  =O\left(  \frac{1}{\gamma^{4}\varepsilon^{2}}\left(  \operatorname*{fat}%
\nolimits_{\mathcal{C}^{\ast}}\left(  \frac{\gamma^{2}\varepsilon}{5}\right)
\log^{2}\frac{1}{\gamma\varepsilon}+\log\frac{1}{\delta}\right)  \right)
\end{align*}
Finally,\ we claim that $\operatorname*{fat}\nolimits_{\mathcal{C}^{\ast}%
}\left(  \eta\right)  \leq2\operatorname*{fat}\nolimits_{\mathcal{C}}\left(
\eta/2\right)  $ for all $\eta>0$. \ To see this, let $\mathcal{C}_{0}$\ be
the p-concept class consisting of $H\left(  x\right)  ^{2}$\ for all
$H\in\mathcal{C}$, and let $\mathcal{C}_{1}$\ be the class consisting of
$\left(  1-H\left(  x\right)  \right)  ^{2}$\ for all $H\in\mathcal{C}$.
\ Then clearly%
\[
\operatorname*{fat}\nolimits_{\mathcal{C}^{\ast}}\left(  \eta\right)
\leq\operatorname*{fat}\nolimits_{\mathcal{C}_{0}}\left(  \eta\right)
+\operatorname*{fat}\nolimits_{\mathcal{C}_{1}}\left(  \eta\right)  .
\]
Also, if $\left\vert H_{1}\left(  x\right)  ^{2}-H_{2}\left(  x\right)
^{2}\right\vert \geq\eta$, then $\left\vert H_{1}\left(  x\right)
-H_{2}\left(  x\right)  \right\vert \geq\eta/2$. \ Hence $\operatorname*{fat}%
\nolimits_{\mathcal{C}_{0}}\left(  \eta\right)  \leq\operatorname*{fat}%
\nolimits_{\mathcal{C}}\left(  \eta/2\right)  $, and similarly
$\operatorname*{fat}\nolimits_{\mathcal{C}_{1}}\left(  \eta\right)
\leq\operatorname*{fat}\nolimits_{\mathcal{C}}\left(  \eta/2\right)  $.
\end{proof}

\begin{proof}
[Proof of Theorem \ref{finelower}]We start with part (i). \ Let
$k=\operatorname*{fine}\nolimits_{\mathcal{C}}\left(  \gamma,\eta\right)  $,
let $S=\left\{  s_{1},\ldots,s_{k}\right\}  $\ be any set that is\ $\left(
\gamma,\eta\right)  $-fine-shattered by $\mathcal{C}$,\ and let $\mathcal{C}%
^{\ast}\subseteq\mathcal{C}$\ be a subclass\ of size $2^{k}$\ that $\left(
\gamma,\eta\right)  $-fine-shatters\ $S$. \ Then the function $F$\ will\ be
chosen uniformly at random from $\mathcal{C}^{\ast}$. \ Also, the distribution
$\mathcal{D}$\ will choose $s_{1}$\ with probability $1-4\varepsilon$, and
otherwise will choose uniformly at random from $\left\{  s_{2},\ldots
,s_{k}\right\}  $. \ Finally,\ the learning algorithm will be given
$p_{i}=\frac{1}{2}-\gamma$ if $F\left(  x_{i}\right)  \leq\frac{1}{2}-\gamma$,
or $p_{i}=\frac{1}{2}+\gamma$ if $F\left(  x_{i}\right)  \geq\frac{1}%
{2}+\gamma$.

First suppose $m\leq\frac{1}{4\varepsilon}\ln\frac{1}{2\delta}$. \ Then the
$m$ samples $x_{1},\ldots,x_{m}$\ will all equal $s_{1}$\ with probability at
least%
\[
\left(  1-4\varepsilon\right)  ^{\frac{1}{4\varepsilon}\ln\frac{1}{2\delta}%
}\geq2\delta.
\]
Conditioned on this happening, the algorithm certainly fails with probability
at least $\frac{1}{2}$.

Next suppose $m\leq\frac{k-1}{64\varepsilon}$. \ Let $r$ be the number of
$i$'s such that $x_{i}\neq s_{1}$. \ Then $\operatorname*{EX}\left[  r\right]
=4\varepsilon m$, and hence%
\[
\Pr\left[  r>\frac{k-1}{4}\right]  \leq\frac{4\varepsilon m}{\left(
k-1\right)  /4}\leq\frac{1}{4}%
\]
by Markov's inequality. \ Furthermore, conditioned on $r\leq\frac{k-1}{4}$,
there are at least $\frac{3}{4}\left(  k-1\right)  $\ indices $i$ for
which\ the algorithm has \textquotedblleft no information\textquotedblright%
\ about $F\left(  x_{i}\right)  $---in other words,\ cannot predict whether
$F\left(  x_{i}\right)  \leq\frac{1}{2}-\gamma$\ or $F\left(  x_{i}\right)
\geq\frac{1}{2}+\gamma$\ better than the outcome of a fair coin toss. \ Yet to
output an $H$ such that%
\[
\Pr_{x\in\mathcal{D}}\left[  \left\vert H\left(  x\right)  -F\left(  x\right)
\right\vert \geq\gamma\right]  \leq\varepsilon<\frac{1}{4},
\]
the algorithm needs to guess correctly for at least $\frac{k-1}{2}$\ of these
$i$'s. \ Again by Markov's inequality, it can do this with probability at most
$\frac{1/2}{3/4}=\frac{2}{3}$. \ Hence it fails with overall probability at
least $\frac{3}{4}\cdot\frac{1}{3}=\frac{1}{4}>\delta$.

Part (ii) can be proved along the same lines as part (i); we merely give a
sketch. \ If $m\leq\frac{1}{4\varepsilon}\ln\frac{1}{2\delta}$, then the
algorithm fails with probability at least $\delta$\ for the same reason as
before. \ Also, as before, the algorithm must be able to guess $F\left(
s_{j}\right)  $\ with probability at least $1-\delta$\ for $\Omega\left(
k-1\right)  $\ indices $j\in\left\{  2,\ldots,k\right\}  $.\ \ But this
requires $m=\Omega\left(  \frac{k-1}{\left(  \gamma+\eta\right)
^{2}\varepsilon}\right)  $\ samples $x_{1},\ldots,x_{m}$. \ For we can think
of each $F\left(  s_{j}\right)  $\ as a coin, whose bias (in the best case) is
either $\frac{1}{2}-\left(  \gamma+\eta\right)  $\ or $\frac{1}{2}+\left(
\gamma+\eta\right)  $. \ And it is known that, to decide\ whether a given coin
has bias $\frac{1}{2}-\left(  \gamma+\eta\right)  $\ or $\frac{1}{2}+\left(
\gamma+\eta\right)  $ with error probability at most $\delta\ll\frac{1}{2}$,
one needs to flip the coin $\Omega\left(  \frac{1}{\left(  \gamma+\eta\right)
^{2}}\right)  $\ times.
\end{proof}

\begin{proof}
[Proof of Theorem \ref{antvthm2}]Follows by small tweaks to the proof of
Theorem \ref{antvthm} found in \cite{antv}. \ Without going into too much
detail, in Ambainis et al.'s construction we first choose a random set
$\mathcal{T}=\left\{  \left(  \pi_{1},r_{1}\right)  ,\ldots,\left(  \pi_{\ell
},r_{\ell}\right)  \right\}  $ of transformations of a certain covering code,
where $\ell=k^{3}$. \ We then use a Chernoff bound to argue that, with high
probability over the choice of $\mathcal{T}$, each of the $k2^{k}%
$\ probabilities $\operatorname*{Tr}\left(  E_{i}\rho_{y}\right)  $\ satisfies
$\left\vert \operatorname*{Tr}\left(  E_{i}\rho_{y}\right)  -q\right\vert
\leq1/k$ for some universal constant $q\geq p+1/k$. \ It follows that there
\textit{exists} a $\mathcal{T}$\ for which this property holds.

Now observe that without loss of generality we can make $q=p+1/k$. \ To do so,
we simply \textquotedblleft dilute\textquotedblright\ each $\rho_{y}$\ to
$c\rho_{y}+\left(  1-c\right)  I$, where $I$ is the maximally mixed state and
$c=\frac{p+1/k-1/2}{q-1/2}$. \ This already proves the theorem\ in the special
case $\eta=2/k$. \ If $\eta<2/k$, then it suffices to repeat Ambainis et al.'s
argument with $\ell=4k/\eta^{2}$ instead of $\ell=k^{3}$.
\end{proof}

\begin{proof}
[Proof of Corollary \ref{antvcor}]It is clear that any lower bound on $k$ that
we can obtain from Theorem \ref{antvthm2}\ is also a lower bound on
$\operatorname*{fine}\nolimits_{\mathcal{C}_{n}}\left(  \gamma,\eta\right)  $.
\ Let $p:=\gamma+\frac{1}{2}$.\ \ Then basic properties of the entropy
function imply that $1-H\left(  p\right)  \leq4\gamma^{2}$. \ Also, let
$k:=\left\lfloor n/5\gamma^{2}\right\rfloor $\ and $\eta:=8\gamma^{2}/n$.
\ Then one can check that the two conditions of Theorem \ref{antvthm2} are
satisfied, as follows. \ Firstly, $k\leq\frac{n}{4\gamma^{2}}=\frac{2}{\eta}$.
\ Secondly,%
\begin{align*}
n &  \geq5\gamma^{2}k\\
&  \geq\left(  1-H\left(  p\right)  \right)  k+\gamma^{2}k\\
&  \geq\left(  1-H\left(  p\right)  \right)  k+\frac{n}{5}-\gamma^{2}\\
&  \geq\left(  1-H\left(  p\right)  \right)  k+\frac{n-5}{5}\\
&  \geq\left(  1-H\left(  p\right)  \right)  k+7\log_{2}\frac{1}{\eta},
\end{align*}
where the last line uses the fact that $\eta\geq2^{-\left(  n-5\right)  /35}$.
\ Hence there exist $n$-qubit mixed states\ $\left\{  \rho_{y}\right\}
_{y\in\left\{  0,1\right\}  ^{k}}$ and measurements $E_{1},\ldots,E_{k}$ such
that for all $y\in\left\{  0,1\right\}  ^{k}$ and $i\in\left\{  1,\ldots
,k\right\}  $:

\begin{enumerate}
\item[(i)] if $y_{i}=0$\ then $\frac{1}{2}-\gamma-\eta\leq\operatorname*{Tr}%
\left(  E_{i}\rho_{y}\right)  \leq\frac{1}{2}-\gamma$, and

\item[(ii)] if $y_{i}=1$\ then $\frac{1}{2}+\gamma\leq\operatorname*{Tr}%
\left(  E_{i}\rho_{y}\right)  \leq\frac{1}{2}+\gamma+\eta$.
\end{enumerate}
\end{proof}

\begin{proof}
[Proof of Theorem \ref{alphathm}]As in the proof of Theorem \ref{alonthm}, let
$\mathcal{D}^{\ast}$\ be the distribution over $\left(  x,b\right)
\in\mathcal{S}\times\left\{  0,1\right\}  $\ obtained by first drawing $x$
from $\mathcal{D}$, then setting $b=1$ with probability $F\left(  x\right)  $.
\ Also, given a hypothesis $H\in\mathcal{C}$, let $H^{\ast}\left(  x,0\right)
=H\left(  x\right)  $\ and $H^{\ast}\left(  x,1\right)  =1-H\left(  x\right)
$. \ Then let $\mathcal{C}^{\ast}$\ be the p-concept class consisting of
$H^{\ast}$\ for all $H\in\mathcal{C}$.

Call $H^{\ast}\in\mathcal{C}^{\ast}$\ an $\alpha$-good function if%
\[
\operatorname*{EX}_{\left(  x,b\right)  \in\mathcal{D}^{\ast}}\left[  H^{\ast
}\left(  x,b\right)  \right]  \leq\alpha+\inf_{C^{\ast}\in\mathcal{C}^{\ast}%
}\operatorname*{EX}_{\left(  x,b\right)  \in\mathcal{D}^{\ast}}\left[
C^{\ast}\left(  x,b\right)  \right]  .
\]
Notice that if $H^{\ast}$\ is $\alpha$-good, then%
\[
\operatorname*{EX}\limits_{x\in\mathcal{D}}\left[  \Delta_{H,F}\left(
x\right)  \right]  \leq\alpha+\inf_{C\in\mathcal{C}}\operatorname*{EX}%
_{x\in\mathcal{D}}\left[  \Delta_{C,F}\left(  x\right)  \right]
\]
as desired.

Now, suppose we are given $m$ samples $\left(  x_{1},b_{1}\right)
,\ldots,\left(  x_{m},b_{m}\right)  $ drawn independently from $\mathcal{D}%
^{\ast}$. \ Then Theorem \ref{occam2} implies that, if we choose $H^{\ast}%
\in\mathcal{C}^{\ast}$\ to minimize%
\[
\sum_{i=1}^{m}H^{\ast}\left(  x_{i},b_{i}\right)  =\sum_{i=1}^{m}\left\vert
H\left(  x_{i}\right)  -b_{i}\right\vert ,
\]
then $H^{\ast}$ will be $\alpha$-good\ with probability at least $1-\delta$,
provided that%
\[
m=O\left(  \frac{1}{\alpha^{2}}\left(  \operatorname*{fat}%
\nolimits_{\mathcal{C}^{\ast}}\left(  \frac{\alpha}{5}\right)  \log^{2}%
\frac{1}{\alpha}+\log\frac{1}{\delta}\right)  \right)  .
\]
Finally,%
\[
\operatorname*{fat}\nolimits_{\mathcal{C}^{\ast}}\left(  \frac{\alpha}%
{5}\right)  \leq\operatorname*{fat}\nolimits_{\mathcal{C}}\left(  \frac
{\alpha}{10}\right)
\]
by the same argument as in Theorem \ref{alonthm}.
\end{proof}

\end{document}